\documentclass[aps,pre,twocolumn,superscriptaddress,showpacs,floatfix]{revtex4-2}
\usepackage{amsmath,amssymb}
\usepackage{amssymb}
\usepackage{mathtools}
\usepackage[pdftex]{graphicx}
\usepackage{epstopdf}
\usepackage{subfigure}
\usepackage[colorinlistoftodos]{todonotes}
\usepackage[maxfloats=50]{morefloats}
\usepackage{hyperref}
\begin{document}

\title{Inherent Structure Landscape of Hard Spheres Confined to Narrow Cylindrical Channels.}

\author{Mahdi Zarif}
\email[]{m\_zarif@sbu.ac.ir}
\affiliation{Department of Physical and Computational Chemistry, Shahid Beheshti University, Tehran 19839-9411, Iran.}

\author{Raymond J. Spiteri}
\affiliation{Department of Computer Science, University of Saskatchewan, Saskatoon, Saskatchewan S7N 5C9, Canada.}

\author{Richard K. Bowles}
\email{richard.bowles@usask.ca}
\affiliation{Department of Chemistry, University of Saskatchewan, SK, S7N 5C9, Canada.}
\affiliation{Centre for Quantum Topology and its Applications (quanTA), University of Saskatchewan, SK S7N 5E6, Canada.}

\begin{abstract}
The inherent structure landscape for a system of hard spheres confined to a hard cylindrical channel, such that spheres can only contact their first and second neighbours, is studied using an analytical model that extends previous results [Phys. Rev. Lett. 115, 025702 (2015)] to provide a comprehensive picture of jammed packings over a range of packing densities. In the model, a packing is described as an arrangement of $k$ helical sections, separated by defects, that have alternating helical twist directions and where all spheres satisfy local jamming constraints. The structure of each helical section is determined by a single helical twist angle, and a jammed packing is obtained by minimizing the length of the channel per particle with respect to the $k$ helical section angles. An analysis of a small system of $N=20$ spheres shows that the basins on the inherent structure landscape associated with these helical arrangements split into a number of distinct jammed states separated by low barriers giving rise to a degree of hierarchical organization. The model accurately predicts the geometric properties of packings generated using the Lubachevsky and Stillinger compression scheme ($N=10^4$) and provides insight into the nature of the probability distribution of helical section lengths.
\end{abstract}


\maketitle

\section{Introduction}
\label{sec:intro}

Hard sphere particle packings play a fundamental role in understanding a broad range of problems in material science. For example, mixtures of hard sphere particles of different sizes give rise to a variety of stable crystal structures~\cite{10.1021/jp804953r,10.1103/physreve.85.021130,10.1103/physreve.79.046714,10.1103/physreve.103.023307}, and amorphous packings are used as the basis for describing the properties of liquids and  glasses~\cite{10.1038/188910a0,10.1063/1.1725362,10.1080/00268979809483148,10.1038/23819,10.1103/physrevlett.84.2064,10.1103/revmodphys.82.789,10.1103/revmodphys.82.2633,10.1063/1.5036657}, as well as colloidal systems~\cite{10.1038/320340a0} and granular materials~\cite{10.1016/0378-4371(89)90034-4,10.1146/annurev-conmatphys-031214-014336}. Sphere packing problems also appear in applications for computer science and information technology~\cite{mezard_2009}. However, rigorous results for sphere packings are difficult to obtain. The most dense jammed packing of single component hard discs in two dimensions (2D) is the triangular packing, with an occupied volume fraction $\phi=\pi/\sqrt{12}$. Recently, it was shown that the face-centred-cubic (FCC) crystal, with $\phi=\pi/\sqrt{18}\approx 0.74$, is the most dense jammed packing of spheres in 3D~\cite{10.4007/annals.2005.162.1065}, but results for amorphous structures are generally obtained using numerical and molecular simulation techniques.

Confining the fluid to small pores or narrow channels limits the number of contacts a particle can make, reducing the type of local structures that can be formed and simplifying the inherent structure landscape~\cite{10.1016/s0378-4371(98)00404-x,2001.11635}.  As a result, it is sometimes possible to obtain analytical results describing the nature of particle jamming. The 2D system of hard discs trapped between two lines, where the confinement only allows contacts up to the second nearest neighbours, has been studied extensively because its local packing environments are highly ordered and the jammed states, from least to most dense, can be characterized in terms of defects~\cite{10.1103/PhysRevE.73.011503,10.1103/PhysRevLett.102.235701}. The comprehensive description of the inherent landscape and the relationship to the properties of the fluid have facilitated a detailed analysis of glassy dynamics~\cite{10.1103/PhysRevLett.109.225701,10.1103/PhysRevLett.110.145701,10.1103/PhysRevE.89.032111,10.1103/PhysRevE.91.022120,10.1103/physreve.91.022301,10.1103/PhysRevE.93.032101}, hyperuniformity in amorphous packings~\cite{10.1103/PhysRevLett.121.075503}, the statistical mechanics of granular materials~\cite{10.1103/PhysRevE.83.031302}, and the possibility of a Gardner transition in hard particle systems~\cite{10.1103/PhysRevLett.120.225501}. Recent studies of this system have also revealed the existence of novel asymptotically crystalline states~\cite{10.1103/physreve.102.042614}.

Both analytical~\cite{10.1103/physrevlett.106.115704,10.1103/physreve.89.042307,10.1063/1.5131318} and simulation~\cite{10.1063/1.2358135,10.1103/physreve.85.051305,10.1103/physreve.79.061111,10.1039/C5SM02875B} studies have shown that spheres confined to quasi-one-dimensional cylindrical channels spontaneously form a variety of structures including, single, double, and staggered helices, as well as some achiral packings, depending on the diameter of the channel~\cite{10.1126/science.181.4101.705,10.1016/0022-5193(80)90290-8,10.1103/physrevlett.85.3652}. These helical structures have also been observed experimentally in molecular~\cite{10.1126/science.1082346,10.1103/PhysRevLett.92.245507}, colloidal~\cite{10.1002/anie.201209767,10.1039/C7SM00316A}, and athermal~\cite{10.1016/j.colsurfa.2014.12.020} systems. Most studies have focused on the formation of the most dense packing with a perfect structure, but to develop connections between the thermodynamics and dynamics of fluids and the inherent structure landscape or to establish a statistical mechanics of athermal systems, it is necessary to understand the nature of defect and amorphous packings in these systems.

The goal of the current work is to develop a comprehensive picture of the inherent structure landscape for a system of hard spheres confined within a narrow cylindrical channel, where particle--particle contacts up to the nearest second neighbours are possible. This builds on our previous analysis~\cite{10.1103/physrevlett.115.025702} by removing the need to use assumptions concerning the nature of the defect states, identifying additional packing environments, and studying the distribution of packings for small systems. As a result, we find that the inherent structure landscape for this system has a degree of hierarchical organization, with basins formed from the arrangements of helical sections splitting into sub-basins, associated with distinct jammed states, separated by small barriers. We also use the model to understand features that appear in the probability distribution of helical section lengths in simulation generated packings.

The remainder of the paper is organized as follows: Section~\ref{sec:model} describes the model under study. Section~\ref{sec:pack} describes the geometry of perfect and defect helical packings. Section~\ref{sec:small} explores inherent structure landscape of the packing model for a small system before the model predictions are compared to the results of large system jammed states produced by simulation in Section~\ref{sec:large}.  Sections~\ref{sec:disc} and~\ref{sec:conc} contain our discussion and conclusions, respectively.  The Appendix outlines the analytical calculation of the random probability distribution for helical section lengths.


\section{Model}
\label{sec:model}

The model studied here consists of \textit{N} three--dimensional hard spheres, with diameter $\sigma$, confined in a cylindrical narrow channel of length \textit{L} with channel diameter
$H_{d}$ in the range of $1+\sqrt{3/4}< H_{d}/\sigma < 2$, which ensures spheres can only contact their first and second neighbours in either direction along the channel. The particle--particle and particle--wall interaction potentials are given by,
\begin{equation}
U(r_{ij})= \left\{\begin{matrix}
0 \;\;\; & r_{ij} \geqslant \sigma \;\;\;\;\;\;\;\;\\
\infty \;\;\; & r_{ij} <  \sigma \;\;\;\;\;\;\;\;
\end{matrix}\right.,
\label{potential}
\end{equation}
\begin {equation}
U_{w}(r_{i})= \left\{\begin{matrix}
0 \;\;\; & |r_{xy}| \leqslant  \left | H_{0}/2 \right | \\
\infty \;\;\; & \textup{otherwise}
\end{matrix}\right.,
\label{wpotential}
\end{equation}           
respectively, where $r_{ij}=\left | \mathbf{r_{j}-r_{i}} \right |$ is the distance between particles, $|r_{xy}|$ is the magnitude of the position vector for a particle perpendicular to the wall where the center of the cylinder is located at $x = y = 0$ and the longitudinal direction of the channel extends in the $z$ direction. The volume accessible to the particles centers is $V_0=\pi L (H_{0}/2)^{2}$, where  $H_{0}=H_{d} - \sigma$, and the occupied volume is $\phi=2 N \sigma ^{3}/\left (3LH^{2}_{d}\right)$.


\section{Geometric Model for Helical Packings}
\label{sec:pack}
\subsection{Perfect Helical Packings}
\label{sec:php}
\begin{figure}[t]
\includegraphics[width=3in]{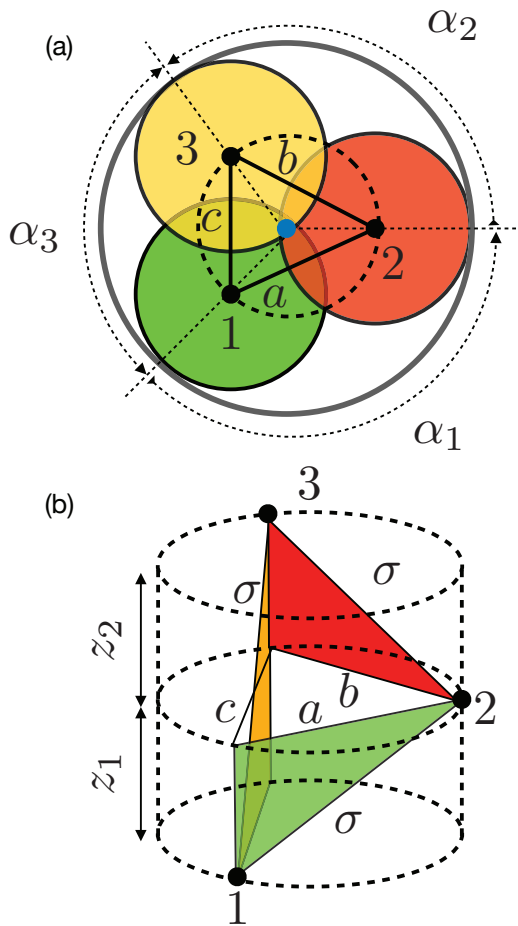}
\caption{Geometric construction for perfect helix. (a) The top view with black points indicating the sphere centers of particles 1, 2, 3 and the blue point marking the central axis of the cylinder. The heavy dotted line outlines the volume accessible to the particle centers and the solid grey line indicates the cylinder wall. (b) The side view showing central accessible volume. The hypotenuses of the colored triangles indicate sphere contacts with lengths $\sigma$. Spheres and external cylinder wall have been remove for clarity.}
\label{fig:geoph}
\end{figure}

This section provides improved details of our geometric analysis presented in Ref.~\cite{10.1103/physrevlett.115.025702}, which focused on the case of $H_{d}/\sigma =1.95$.
The most dense packing of the current system is a helix~\cite{10.1103/physrevlett.85.3652} where each sphere has four sphere-sphere contacts, formed with its first and second neighbours and a single sphere wall contact. Figure~\ref{fig:geoph} shows the geometric construction used to determine the properties of the helix. Particle one is placed in the channel at a fixed point, then particle two is placed along the channel at a  distance $z_1$ and angle $\alpha_1$ in a anticlockwise direction looking from the top, such that it contacts sphere one and the wall. Particle three, which is placed at a distance $z_2$ and $\alpha_2$ along the channel is then constrained to contact the wall as well as particles one and two. Subsequent particles are added, alternating between increments of $(z_1,\alpha_1)$ and $(z_2,\alpha_2)$, maintaining the contact constraints until particles $N-1$ and $N$ are forced to contact particles one and two to ensure that the helical periodic boundary conditions are enforced.

Using this geometric construction yields the following relationships:
\begin{equation}
\sigma^{2} = a^{2} + z_{1}^{2} \mbox{,}
\label{eq:3D_sigma_a}
\end{equation}
\begin{equation}
\sigma^{2} = b^{2} + z_{2}^{2},
\label{eq:3D_sigma_b}
\end{equation}
\begin{equation}
\sigma^{2} = c^{2} + (z_{1}+z_{2})^{2} \mbox{,}
\label{eq:3D_sigma_c}
\end{equation}
\begin{equation}
a = H_{0}\sin (\alpha_{1}/2) \mbox{,}
\label{eq:3D_a}
\end{equation}
\begin{equation}
b = H_{0}\sin (\alpha_{2}/2) \mbox{,}
\label{eq:3D_b}
\end{equation}
\begin{equation}
c = H_{0}\sin (\alpha_{3}/2)\mbox{,}
\label{eq:3D_c} 
\end{equation}
where $a$, $b$, and $c$ form the projected triangle connecting the particle centers. Expressions for $z_1$ and $z_2$ can be obtained by using Eqs.~\ref{eq:3D_a} and \ref{eq:3D_b} in Eqs.~\ref{eq:3D_sigma_a} and \ref{eq:3D_sigma_b} to give respectively,
\begin{equation}
z^{2}_{1} + \frac{H_{0}^{2}}{2} \left[ 1 - \cos \alpha_{1}\right] = \sigma^{2}
\label{eq:3D_Z1}
\end{equation}
and
\begin{equation}
z^{2}_{2} + \frac{H_{0}^{2}}{2} \left[ 1 - \cos \alpha_{2}\right] = \sigma^{2}\mbox{.}
\label{eq:3D_Z2}
\end{equation}
Substituting these results into Eq.~\ref{eq:3D_sigma_c} and using the constraint, $\alpha_{1}+\alpha_{2}+\alpha_{3} = 2\pi$ in Eq.~\ref{eq:3D_c} gives,
\begin{multline}
\sigma^{2} =\left( \sqrt{\sigma^{2} - \frac{H_{0}^{2}}{2} + \frac{H_{0}^{2}}{2} \cos \left[\frac{\alpha_{1}}{2} \right]}\right.\\
+\left. \sqrt{\sigma^{2} - \frac{H_{0}^{2}}{2} + \frac{H_{0}^{2}}{2} \cos \left[\frac{\alpha_{2}}{2} \right]} \right)^{2}\\
- \frac{H_{0}^{2}}{2} \left( \cos \left[\alpha_{1} +\alpha_{2} \right] - 1 \right) \mbox{,}
\label{eq:3D_sigma}
\end{multline}
which can be solved numerically to provide values of $\alpha_2$ as a function of $\alpha_1$. 

\begin{figure}[t]
\includegraphics[]{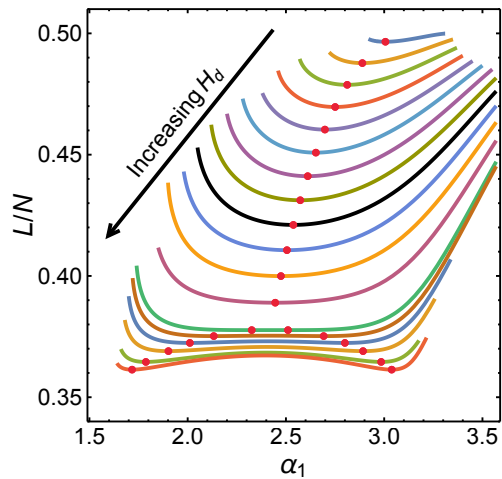}
\caption{Length per particle, $L/N$, as a function of $\alpha_1$, for perfect helical sphere packings with channels diameters in the range $H_d/\sigma=1.87-1.999$. Red points indicate local minima associated with jammed packings.}
\label{fig:lon}
\end{figure}

All the spheres in the helix satisfy the three-dimensional local jamming condition, where each particle has at least four contacts that are not all contained within the same hemisphere. However, this does not guarantee the system is collectively jammed because the concerted motion of particles can lead to unjamming~\cite{10.1021/jp011960q}. In the limit that $H_{d}/\sigma \rightarrow 1+\sqrt{3/4}$, Eq.~\ref{eq:3D_sigma} has a single solution, with $\alpha_{1} = \alpha_{2} = \pi$, that corresponds to the formation of the expected single zig--zag chain of jammed particles. For wider channels, there is a continuous range of solutions where the variation of $\alpha_1$ produces a concerted twisting motion of the helix that compresses the overall structure. To locate the most dense jammed state of the system, we then minimize the length per particle,
\begin{equation}
\frac{L}{N}=\frac{1}{2}\left(z_{1} + z_{2}\right) \mbox{.}
\label{eq:3Dlon}
\end{equation}
Figure~\ref{fig:lon} shows that $L/N$ exhibits a single minimum as a function of $\alpha_1$ in the range $1+\sqrt{3/4}<H_d/\sigma<1+4\sqrt{3}/7$, where $\alpha_1=\alpha_2$ and $z_1=z_2$, confirming that the most dense packings are single helices. In the region $1+4\sqrt{3}/7<H_d/\sigma<2$, $L/N$ $\alpha_1$ no longer equals $\alpha_2$ and, the resulting jammed structure consists of two staggered helices, which leads to the appearance of two identical minima corresponding to the two possible ways of alternating between  $\alpha_1$ or $\alpha_2$. It is also important to note that both the right ($\mathcal{P}$) and left ($\mathcal{M}$) handed helices can be constructed by incrementing the angle in clockwise or anticlockwise directions respectively. The results from our geometric construction are consistent with the simulation results obtained by Picket et al.~\cite{10.1103/physrevlett.85.3652}. It is also important to note that Fig.~\ref{fig:lon} essentially reproduces the results of Chen et al.~\cite{10.1063/1.5131318}, who used  a similar approach to study the most dense perfect helical packings for $1+\sqrt{3/4}< H_{d}/\sigma < 2$, and we include this figure here for completeness and to highlight the importance of the minimization of $L/N$ with respect $\alpha_1$, which plays an expanded role in determining the nature of jammed structures containing defects.

\subsection{Defect Helical Packings}
\label{sec:dhp}

\begin{figure}[ht]
\includegraphics[width=3in]{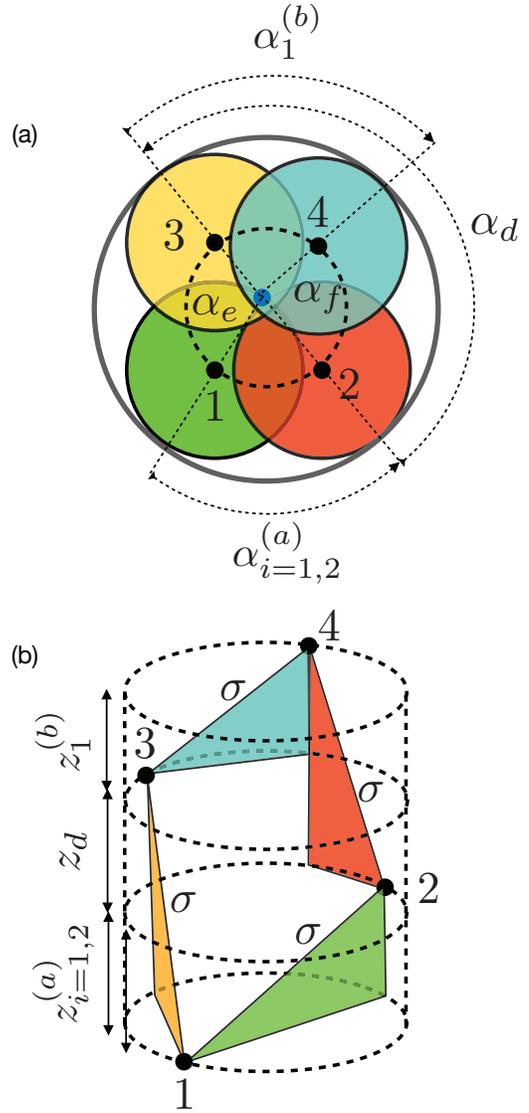}
\caption{Geometric construction for helical defects (a) The top view, and (b) the side view.}
\label{fig:geodef}
\end{figure}

To generate lower density jammed states, we introduce topological defects into the helix that reverse the direction of the helical twist, noting that these must be introduced in pairs in order to maintain the helical periodic boundary conditions. As a result, a packing with $k$ defects consists of $k$ alternating left and right twisting sections of helix that can be characterized by a list $\left\{n^{(1)},n^{(2)},\ldots,n^{(k)}\right\}$, where $n^{(j)}$ specifies the number of spheres in a helical section $j$ so that $N=\sum_{j=1}^k n^{(j)}$. Figure~\ref{fig:geodef} shows the geometric construction used to calculate the properties of a defect located between two sections of helix, denoted with superscripts $a$ and $b$ respectively. Spheres 1 and 2, which are in contact, represent the last two particles of helix section $a$ twisting in an anti-clockwise direction (from the top) and the angle $\alpha^a_i$ can be either $\alpha_1$ ($i=1$) if $n^{(a)}$ is even or $\alpha_2$ ($i=2$) if $n^{(a)}$ is odd. Spheres 3 and 4 are also in contact with each other and are the first two particles in helix section $b$, twisting in a clockwise direction. The defect is then located between particles 2 and 3 and is characterized by the longitudinal length $z_d$ and angle $\alpha_d$. In order to satisfy local jamming constraints, all the particles must contact the walls, sphere 3 must also contact sphere 1, and sphere 2 must contact sphere 4, but a contact between spheres 2 and 3 is not necessary. 

Using the geometric construction shown in Fig.~\ref{fig:geodef}, we obtain the following relations for the particle separations along the longitudinal $z$-axis:
\begin{equation}
(z^{(a)}_{i})^2 + \frac{H_{0}^{2}}{2} \left[ 1 - \cos \alpha^{(a)}_{i}\right] = \sigma^{2}\mbox{,}
\label{eq:za1}
\end{equation}
\begin{equation}
(z^{(b)}_{1})^2 + \frac{H_{0}^{2}}{2} \left[ 1 - \cos \alpha^{(b)}_{1}\right] = \sigma^{2}\mbox{,}
\label{eq:zb1}
\end{equation}
\begin{equation}
(z_{e})^2 + \frac{H_{0}^{2}}{2} \left[ 1 - \cos \alpha_{e}\right] = \sigma^{2}\mbox{,}
\label{eq:ze}
\end{equation}
\begin{equation}
(z_{f})^2 + \frac{H_{0}^{2}}{2} \left[ 1 - \cos \alpha_{f}\right] = \sigma^{2}\mbox{,}
\label{eq:zf}
\end{equation}
where $\alpha_{e}$ and $\alpha_{f}$ are the in-plane angles between particles one and three and particles two and four, respectively. Substituting Eqs.~\ref{eq:za1}--\ref{eq:zf}, along with the angular constraints $\alpha_e=2\pi-\alpha^{(a)}_i-\alpha_d$ and $\alpha_f=\alpha_d-\alpha^{(b)}_1$ into  $(z_e)^2=(z^{(a)}_{i}+z_d)^2$ and $(z_f)^2=(z^{(b)}_{1}+z_d)^2$ yields two relations, 
\begin{multline}
z_d^2+ \frac{H_{0}^{2}}{2}\cos \alpha^{(a)}_i+2 z_d\sqrt{\sigma^2-\frac{H_{0}^{2}}{2}\left[1-\cos\alpha^{(a)}_i\right]}\\
-\frac{H_{0}^{2}}{2}\left[\cos(\alpha^{(a)}_i+\alpha_d)\right]=0\mbox{,}
\label{eq:zd}
\end{multline}
and
\begin{multline}
\frac{H_{0}^{2}}{2}\left[1-\cos(\alpha_d-\alpha_f)\right]-\sigma^2\\
+\left(z_d+\sqrt{\sigma^2-\frac{H_{0}^{2}}{2}\left[1-\cos\alpha^{(b)}_1\right]}\right)^2=0\mbox{,}
\label{eq:alphad}
\end{multline}
that can be solved numerically to obtain $z_d$ and $\alpha_d$. Equations~\ref{eq:zd} and \ref{eq:alphad} highlight the fact that the geometric properties of the defect are a function of the properties of the helical sections on either side through their dependence on $\alpha^{(a)}_i$ and $\alpha^{(b)}_1$. This represents an improvement on our original analysis in Ref.~\cite{10.1103/physrevlett.115.025702}, where it was assumed $\alpha_d=\pi$, a condition that is only true if the helices on both sides of the defect are the same, i.e., for what we referred to as helical defect crystals because they represent repeated units of identical left-- and right--handed helical sections so the defects appear at regular spacings along the packing.

The length per particle of a configuration of spheres satisfying the contact conditions and containing $k$ defects is then given by,
\begin{multline}
\frac{L}{N}=\frac{1}{N}\sum_{j=1}^k\left[ \left(\frac{n^{(j)}-A}{2}\right)z_1(\alpha_1^{(j)})\right.\\
+\left.\left(\frac{n^{(j)}-A}{2}-B\right)z_2(\alpha_1^{(j)})+z_d(\alpha_i^{(j)},\alpha_1^{(j+1)})\right]\mbox{,}
\label{eq:deflon}
\end{multline}
where $A=0\mbox{ or }1$ if $n^{(j)}$ is even or odd respectively, and $B=1\mbox{ or }0$ if $n^{(j)}$ is even or odd respectively. Helical periodic boundaries are imposed for the first and last helical sections, noting that in determining $z_d$, $i=1$ for even $n^{(j)}$ and $i=2$ for odd $n^{(j)}$. A jammed packing can then be obtained by minimizing $L/N$ in Eq.~\ref{eq:deflon} with respect to the set of $k$ angles, $\alpha_1^{(j)}$, that characterize each helical section and the jamming density, $\phi_J=2 N \sigma ^{3}/\left ( 3LH^{2}_{d}\right)$.

The number of possible helical arrangements for a given defect fraction, $\theta=k/N$, can be found by considering the number of ways we can distribute $N$ indistinguishable spheres into $k$ helical sections. The geometric contact restrictions of the defects require each section of helix to contain at least two spheres, which leaves $N-2k$ spheres to be placed without restrictions, yielding the number of helical section arrangements with $k$ defects,
\begin{equation}
N_h(k)=\frac{2(N-k-1)!}{(k-1)!(N-2k)!}\mbox{,}\\
\label{eq:njk}
\end{equation}
where the factor for $2$ appears because we can choose to begin twisting the first helical section in either a clockwise or anticlockwise direction. Taking the natural logarithm of Eq.~\ref{eq:njk} and using Stirling's approximation then gives, in the thermodynamic limit,
\begin{equation}
\frac{\ln N_h(\theta)}{N}=(1-\theta)\ln(1-\theta)-\theta\ln\theta-(1-2\theta)\ln(1-2\theta)\mbox{,}\\
\label{eq:lnnj}
\end{equation}
where $0 \le \theta \le 0.5$. The total number of  helical arrangements can be obtained by summing over all possible values of $k=0,1,\ldots,N/2$ in Eq.~\ref{eq:njk}, to obtain,
\begin{equation}
\lim_{N\rightarrow\infty}\frac{\ln N_h}{N}=\ln\left(\frac{1+\sqrt{5}}{2}\right)\approx 0.481\mbox{.}\\
\label{eq:njtotal}
\end{equation}
Equation~\ref{eq:lnnj} is the same distribution of structures found in the 2D model of hard discs confined between two lines where only nearest neighbour contacts are allowed~\cite{10.1103/PhysRevE.73.011503}. However, in the 2D model, $\theta$ is directly linked to the jamming density, $\phi_J$, which is independent of the way the defects are arranged, and there is only one jammed state associated with each defect arrangement. As we will see, this is not the case for the current model, so $N_h(\theta)$ becomes a lower bound on the number of jammed states, and $\phi_J$ for a packing is not only determined by $\theta$, but it also depends on how the defects are distributed.

\section{Small System Jammed Packings}
\label{sec:small}

\begin{figure}[b]
\includegraphics[]{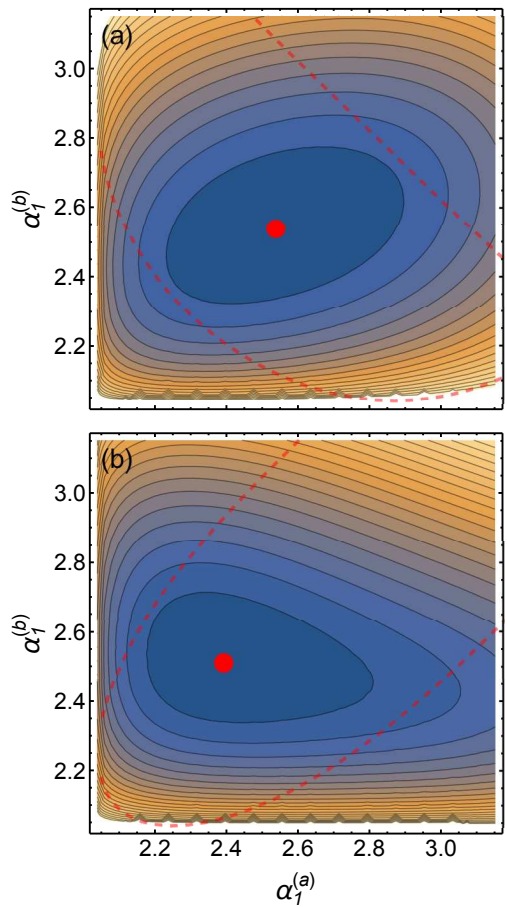}
\caption{Contour plots of $L/N$ as a function of $\alpha_{1}^{(a)}$  and $\alpha_{1}^{(b)}$ for packings (a) $\{15,35\}$ and (b) $\{14,36\}$, with $H_d/\sigma=1.95$. Contours are separated by 0.001, the red dashed lines represent the contact constraint between particles 2 and 3 in the defects, and the red points denote the minima.}
\label{fig:oddeven1}
\end{figure}

To understand how the presence of defects affects the overall structure of the packings, we begin by studying a system of $N=50$ spheres containing just two defects and two sections of helix, denoted $a$ and $b$, for a case where $H_d/\sigma=1.95$. Figure~\ref{fig:oddeven1}a shows a contour plot of $L/N$, from Eq.~\ref{eq:deflon}, as a function of $\alpha^{(a)}_1$ and $\alpha^{(b)}_1$ for a system $\{n^{(a)},n^{(b)}\}=\{15,35\}$. When the defects are well separated and both helix sections contain an odd number of spheres, the minimum occurs where $\alpha^{(a)}_1=\alpha^{(b)}_1$, at a value where $\alpha_1=\alpha_2$ within both sections of helix. As a result, the jammed packing consists of two sections of a perfect single helix, identical to the most dense structure, with opposite $\mathcal{P}$ and $\mathcal{M}$ twists, separated by a large volume defect.

\begin{figure}[t]
\includegraphics[]{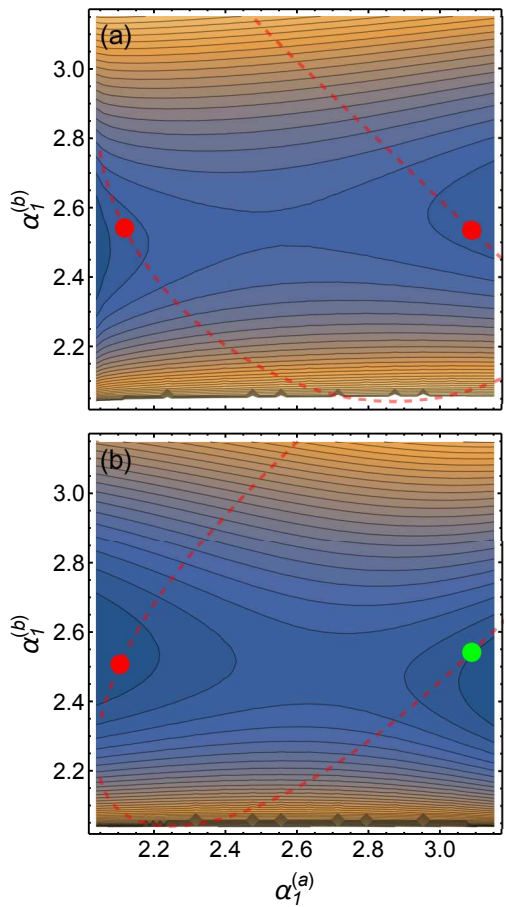}
\caption{Contour plots of $L/N$ as a function of $\alpha_{1}^{(a)}$  and $\alpha_{1}^{(b)}$ for packings (a) \{3,47\} and (b) \{2,48\}, with $H_d/\sigma=1.95$. Contours are separated by 0.001, the red dashed lines represent the contact constraint between particles 2 and 3 in the defects and the red and green points denote a global minimum and a local minimum, respectively.}
\label{fig:oddeven2}
\end{figure}

When the two helical sections contain an even number of spheres, such as $\{n^{(a)},n^{(b)}\}=\{14,36\}$ (see Fig.~\ref{fig:oddeven1}b), $\alpha^{(a)}_1\neq\alpha^{(b)}_1$ so that the two sections no longer have the same structure. Furthermore, within each section, $\alpha_1\neq\alpha_2$, indicating the structure has changed to that of a double helix. In this case, the presence of a single defect has disrupted the entire global structure of the packing. Figure~\ref{fig:oddeven1} also shows that  when the defects are well separated (large helical sections) the minima are located away from the contact constraint between particles 2 and 3 in the defect indicating there is a gap between the particles. As the defects are brought closer together, by decreasing the size of one of the helical sections, the surface of the minimum elongates along the $\alpha^{(a)}_1$ axis before eventually splitting in two when $n^{(a)}\approx10$. Figure~\ref{fig:oddeven2} shows that the system becomes jammed when particles 2 and 3 in the defect come into contact, locking the structure in place. Now, $\alpha^{(a)}_1\neq\alpha^{(b)}_1$ for odd--numbered helical sections, but the two minima have the same $L/N$ and result from a switching between the values of $\alpha_1$ and $\alpha_2$ within a section. Nevertheless, the two packings are distinct structures because they are separated by a barrier, and it requires the concerted motion of the relative position of all the particles to move from one state to the other. The jamming densities of the two states are not the same for the even--number helical sections (Fig.~\ref{fig:oddeven2}).

\begin{figure}[t]
\includegraphics[]{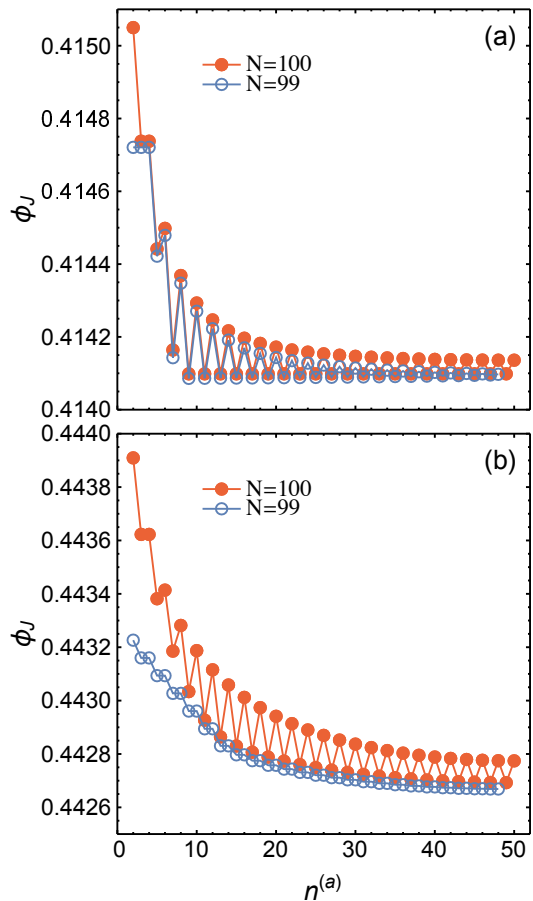}
\caption{The jamming occupied volume fraction, $\phi_J$, for packings $\{n^{(a)},N-n^{(a)}\}$ as a function of $n^{(a)}$ for systems with $N=100$ (filled circles) and $N=99$ (open circles) for (a) $H_d/\sigma=1.95$ (b) $H_d/\sigma=1.99$. Only the most dense structure for each $\{n^{(a)},N-n^{(a)}\}$ is included.}
\label{fig:ddinteration}
\end{figure}

Figure~\ref{fig:ddinteration}a shows $\phi_J$, where Eq.~\ref{eq:deflon} has been minimized, as a function of $n^{(a)}$ for a system containing two defects, with $H_d/\sigma=1.95$, and we have only identified the global density maxima ($L/N$ global minima) for each $n^{(a)}$. For $N=100$, as $n^{(a)}$ is varied, the structure $\{n^{(a)},N-n^{(a)}\}$ oscillates between structures where both sections of helix are odd-sized and then both sections are even-sized. The jamming density of the even-sized sections increases as $n^{(a)}$ decreases, bringing the defects closer together, whereas $\phi_J$ for the structures with odd-sized sections remains constant until $n^{(a)} <10$. When $N=99$, $n^{(a)}$, the smaller of the two helical sections, oscillates between odd and even, and we see a similar trend in $\phi_J$, where $\phi_J$ is lower when $n^{(a)}$ is odd. Figure~\ref{fig:ddinteration}b shows the same analysis for a system with $H_d/\sigma=1.99$, where the most dense state is a double helix. The behaviour of $\phi_J$ for the even helical sections remains the same as observed for the narrower channel and increases as $n^{(a)}$ decreases. However, we now see $\phi_J$ for the odd helical sections increasing over the entire region. In addition, $\alpha_i$ in both odd helical sections differ from those of the perfect helix.

The analysis of a system with just two defects clearly shows that the way the defects are distributed within a packing influences $\phi_J$, suggesting it is worth examining packings with an increased number of defects. The total number of jammed packings increases exponentially with $N$, but the simplicity of the packing model described in Section~\ref{sec:pack} means that it is possible to generate all the possible model packings for small systems and examine the full jamming landscape. The challenge of finding all the jammed structures is simplified by recognizing that all the helical arrangements, under periodic boundaries, can be generated from a series of canonical arrangements by a rotation, $\{n^{(1)},n^{(2)},\ldots,n^{(k)}\}\rightarrow\{n^{(k)},n^{(1)},\ldots,n^{(k-1)}\}$ or a reversal of the order, $\{n^{(1)},n^{(2)},\ldots,n^{(k)}\}\rightarrow\{n^{(k)},n^{(k-1)},\ldots, n^{(1)}\}$, then any replicated states can be removed. For example, the canonical arrangement $\{2,2,3,2,2,9\}$, generates six distinct arrangements by rotation, but the operation of reversal generates six identical arrangements, which are then removed to leave the original six. However, for $\{2,2,2,3,2,9\}$, the rotation and reversal operations generate a total of 12 distinct arrangements.

All structures related to the same canonical arrangement maintain the same relative position of all the helical sections and minimize to a set of jammed structures with the same respective geometric properties, with the same set of $\alpha^{(j)}_1$ and jamming density. In addition, each helical arrangement actually represents two possible arrangements that can be formed by winding in the opposite direction, but the arrangements will again minimize to structures with the same geometric properties. This mechanical approach lets us count the number of states associated with each canonical arrangement and reproduces the same total number of helical arrangements as predicted by Eq.~\ref{eq:njk}. Table~\ref{tab:con} compares the number of canonical arrangements, $N_c$, with the total number of helical arrangements, $N_h$, for a series of small system sizes and highlights the significant reduction in the number of arrangements that need to be analysed.

\begin{table}[b]
\caption{The number of canonical helical arrangements compared to the total number for small systems. }
\begin{center}
\begin{tabular}{rrr}
\hline
\hline
$N$ & $N_h$ & $N_c$ \\
\hline
16 & 306 & 54 \\ 
20 & 2091 & 238\\
24 & 14330 & 1206 \\
32 & 673135 & 38889\\
\hline
\hline
\end{tabular}
\end{center}
\label{tab:con}
\end{table}%

 Here, we study systems of $N=20$ spheres with $H_d/\sigma=1.95$ and $1.99$, respectively, by searching for the jammed packings associated with all of their 238 canonical helical arrangements. Since the basin for a given helical arrangement can split into sub--basins, as shown in Fig.~\ref{fig:oddeven2}, we use the local optimization, Nelder-Mead algorithm in Mathematica~\cite{mv12} and perform five minimizations of Eq.~\ref{eq:deflon} for each helical arrangement with randomized initial starting points. While this does not ensure we find the global minimum for each helical arrangement, and a larger number of trials for each arrangement could reveal additional jammed states, it allows us to explore the general features affecting the appearance of sub--basins. We will also focus our analysis on arrangements that start with a single direction of twist for the first helical section as the other starting twist direction will yield the same results for $\phi_J$. 

To begin, it is useful to examine some of the properties of a few particular packings. The perfect helix arrangement, $\{20\}$, i.e., with $\theta=0$, has a single canonical structure that leads to one and two jammed structures for $H_d/\sigma=1.95$ and $1.99$, respectively, as discussed earlier. The $\{10,10\}$ canonical arrangement represents a single arrangement that minimizes to a single jammed structure, where $\alpha^j_1$ is the same for all helical sections, which is also true for $\{2,2,2,2,2,2,2,2,2,2\}$ ($\theta=0.5$).  For the $\{5,5,5,5\}$ arrangement, we see a more complex set of jammed states. With $H_d/\sigma=1.95$, we find two distinct packings where $\alpha^{(j)}_1$ is the same in all four helical sections, one where the list of angles $\{\alpha^{(j)}_1\}=\{2.2943,2.2943,2.2943,2.2943\}$ and the other with $\{\alpha^{(j)}_1\}=\{2.8104,2.8104,2.8104,2.8104\}$. We can understand the appearance of these packings by noting the helical sections containing an odd number of spheres have an even number of alternating angles, $\alpha^{(j)}_1$ and $\alpha^{(j)}_2$ (see Eq.~\ref{eq:deflon}), where the second angle is a function of the first. In the first packing, $\alpha^{(j)}_1=2.2943$ and $\alpha^{(j)}_2=2.8104$, so it is possible to construct the second packing by simply interchanging the two angles. For $H_d/\sigma=1.99$, these uniform packings,  where $\alpha^{(j)}_1$ is the same in all helical sections, are replaced by packings that alternate $\alpha^{(j)}_1$; i.e., $\{\alpha^{(j)}_1\}=\{1.7639,2.7512,1.7639,2.7512\}$ and $\{\alpha^{(j)}_1\}=\{2.7512,1.7639,2.7512,1.7639\}$. In addition, we also find that the $\{5,5,5,5\}$ helical arrangement, for both channel diameters, jams in a series of asymmetrical structures, where two of the helical sections are the same, but the remaining two are different; i.e., for $H_d/\sigma$=1.95 we find $\{\alpha^{(j)}_1\}=\{2.5378,3.0837,2.5378,2.1206\}$, which functions as a canonical structure for four additional jammed states. As a result, the inherent structure basin associated with the $\{5,5,5,5\}$ helical arrangement splits into a total of six distinct jammed states. These sets of structures illustrate the complexity of the packings that arise, despite the simplicity of the model, and show that while Eqs.~\ref{eq:njk}--\ref{eq:njtotal} provide a count of the number of basins, or meta--basins, they represent a lower bound to the number of jammed states. Most of the remaining canonical structures minimize to at least two distinct jammed states, and it is possible additional structures would be found with a more extensive search.

\begin{figure}[t]
\includegraphics[]{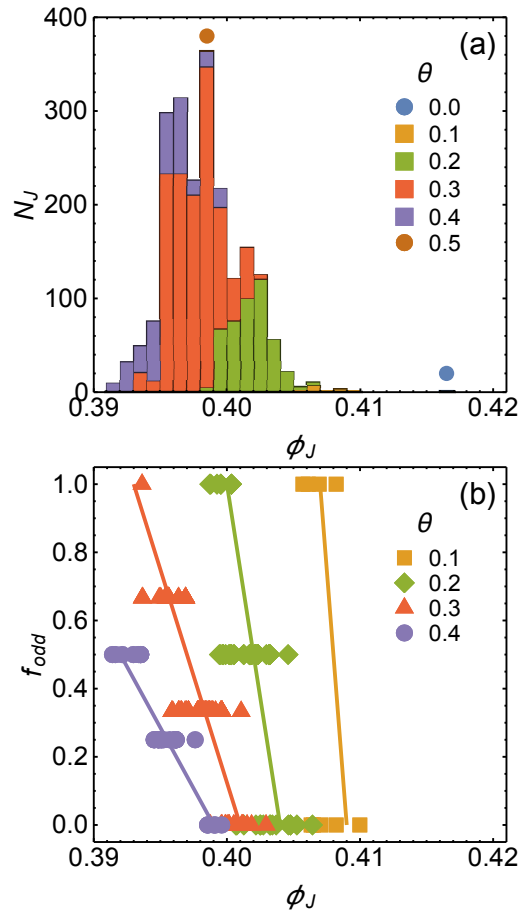}
\caption{(a) The cumulative number distribution of jammed states, $N_J$, with different $\theta$, as a function of $\phi_J$, for $H_d/\sigma=1.95$.  The $\phi_J$ for the  jammed structure with $\theta=0$ and $\theta=0.5$ are highlighted with blue and brown circles, respectively. The bin size for the distribution $\Delta\phi_J=0.001$. (b) Fraction of odd helical sections, $f_{odd}$, for each $\theta$ from the same distribution. Solid lines provide a guide to the eye.}
\label{fig:njdist}
\end{figure}

A number of factors affect the $\phi_J$ of a structure, including the number of defects and how the defects are distributed. To capture some of the more generic features of the jammed states, we plot (Fig.~\ref{fig:njdist}a) the distribution of most dense jammed states obtained for each canonical structure, weighted by the number of helical arrangements it represents, which gives rise to a distribution of 2091 structures for $N=20$, for the $H_d/\sigma=1.95$ model. Similar results were obtained for the $H_d/\sigma=1.99$ case (see the Supplemental Material (SM)~\cite{supmat}) . Each defect fraction, $\theta$, exhibits a broad distribution of $\phi_J$, again indicating that the jamming density is sensitive to the way the defects are organized and not just to their concentration. Notably, there is a significant gap between the $\phi_J$ of the most dense packing of the perfect helix and the packings with a single defect, but this is the result of the small system size and the distribution fills the density gap as $N$ increases. Figure~\ref{fig:njdist}a also shows that the $\theta=0.5$ jammed structure, which is the maximum defect fraction possible in the model, is not the least dense structure. Structurally, this packing is made up of two parallel zig-zag chains~\cite{10.1103/physrevlett.115.025702}. It becomes the least dense packing in the limit $H_d/\sigma \rightarrow 1+\sqrt{3/4}$, where the particles can only touch their first nearest neighbours, but as $H_d/\sigma\rightarrow 2$,  $\phi_J(\theta=0.5)\rightarrow\phi_J(\theta=0)$ and the two structures, ($\theta=0.5;\theta=0$) become equivalent, leading to the formation of an achiral most dense packing of alternating ``doublets"~\cite{10.1103/physrevlett.85.3652}. However, the full distribution of jammed states is not contained between these two apparent limits, and the $\theta=0.5$ structure moves through the distribution as the channel becomes wider. 

The underlying double helix structure of the packings means that sections of helix containing odd and even numbers of spheres have different jamming properties. To examine this effect in general, we calculate the fraction of helical sections in a jammed state containing an odd number of spheres, $f_{odd}$, for each $\theta$, as a function of $\phi_J$. Figure~\ref{fig:njdist}b shows that increasing $f_{odd}$ tends to decrease the jamming density, with the effect becoming greater for larger $\theta$. This suggests that smaller sections, with odd numbers of particles, tend to pack less efficiently because they generate large defects. For the current small system studied here, the lowest density packings are formed from structures containing a mixture of helical sections with two and three particles.

\section{Large System Simulated Jammed Packings}
\label{sec:large}
\begin{figure}[t]
\includegraphics[]{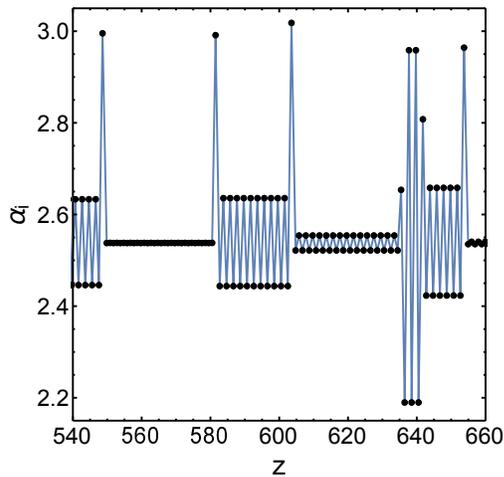}
\caption{The cylindrical angle between neighbouring spheres, $\alpha_{i}$, as a function of the position of the sphere along the $z-$axis of the channel.}
\label{fig:random}
\end{figure}

The model developed here suggests that the jammed packings of this system adopt a ``poly-helical" type structure, containing a mixture of well formed sections of single or double helix, where the properties of each helical section, such as helical pitch, depend on the distribution of the defects.  To test this, we generate jammed packings by continually compressing a system of $N=10^4$ spheres from low density, $\phi=0.01$, using a modified version of the Lubachevsky and Stillinger (LS) molecular dynamics (MD) scheme~\cite{10.1007/bf01025983}, with compression rates $d\sigma/dt=1\times10^{-3}$, $5\times10^{-4}$, and $5\times 10^{-5}$ in reduced units. Packings formed in this way have previously been shown to be jammed and follow the free volume equation of state near their jamming densities, with slower compression rates leading to increased $\phi_J$, and decreased $\theta$~\cite{10.1103/physrevlett.115.025702}. The average properties of the packing generated for this study are listed in Table~\ref{tab:packing}. Figure~\ref{fig:random} shows the cylindrical angle, $\alpha_i$, between successive spheres, as a function of sphere position along the channel, for a typical section of a jammed packing. The well formed sections of double helix can be clearly identified, with their alternating $\alpha_1$, $\alpha_2$ angles, as can the presence of the defects separating the helical sections. The configuration also shows the presence of a section of single helix where $\alpha_1=\alpha_2$, containing an odd number of particles, as predicted by the model.

\begin{table}[b]
\caption{Average properties of simulation generated packings, $N=10000$ and $H_d/\sigma=1.95$, using compression rate $d\sigma/dt$. The standard deviation in the last digit is given in brackets.}
\begin{center}
\begin{ruledtabular}
\begin{tabular}{cccc}
$d\sigma/dt$ & $n^{\circ}$ packings & $\left<\phi_J\right>$ &  $\left<\theta\right>$ \\ 
\hline
$1\times10^{-3}$ & 65 & 0.4117(1)  &  0.044(1) \\ 
$5\times10^{-4}$ & 65 & 0.4129(1)  & 0.033(1) \\
$5\times10^{-5}$ & 153 & 0.4151(2)  & 0.011(2)\\
\end{tabular}
\end{ruledtabular}
\end{center}
\label{tab:packing}
\end{table}

\begin{figure}[t]
\includegraphics[]{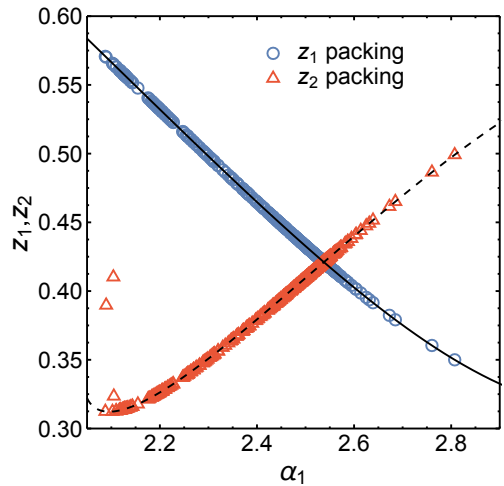}
\caption{Geometric properties, $z_1$ (circles) and $z_2$ (triangles), of helical sections in a jammed packing, formed using  $d\sigma/dt=0.001$ as a function of measured $\alpha_1$, compared to predictions of the model (solid and dashed lines respectively).}
\label{fig:compz1}
\end{figure}

As a check on the validity of the packing model, we compare the geometric parameters, $z_1$, $z_2$, and $\alpha_1$, obtained from the simulation, for each helical section, with the predictions of Eqs.~\ref{eq:3D_Z1}--\ref{eq:3D_sigma}. For almost all the spheres in the generated packings, we find $\Delta z_i=z_i-z_i^{\prime}<5.0\times10^{-6}$, where $z_i$ $(i=1,2,d)$ is measured from the configurations obtained by simulation and $z_i^{\prime}$ is calculated from the model using the measured $\alpha_i$. However, we do find configurations, (Fig.~\ref{fig:compz1}), that show the presence of  particles adopting spacings not predicted by the model at small values of $\alpha_1$, which suggests there may be some packing environments not included in the model.

\begin{figure*}[t]
\includegraphics[]{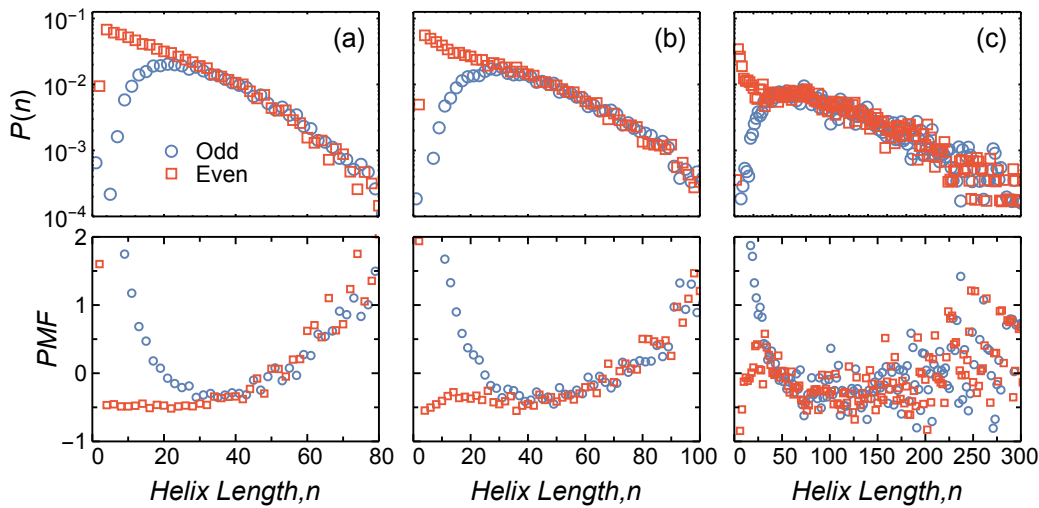}
\caption{(Top Row) Probability, $P(n)$, of finding a helix section containing $n$ spheres for jammed packing with $H_d=1.95$ and $N=10000$. formed with (a) $d\sigma/dt=1\times 10^{-3}$, (b) $d\sigma/dt=5\times 10^{-4}$ and (c) $d\sigma/dt=5\times 10^{-5}$. (Bottow Row) Corresponding potential of mean force.}
\label{fig:helixdist}
\end{figure*}

The packings can be further characterized in terms of the probability, $P(n)$, of finding a helical section of length $n$, where we calculate the distribution over packings formed with the same compression rate because they have similar $\theta$ and $\phi_J$. The location of the defects in the simulation generated packings can be identified using the local geometric properties of the particles  (see SM~\cite{supmat} for details), which includes calculating a volume, $v_{tet}(i)$ for each particle as the volume of the tetrahedron formed by particles, $i-1$, $i$, $i+1$, and $i+2$. The sign of $v_{tet}(i)$ captures the local twist direction of the helix so the particles in the same helical section have the same sign, and in most cases, the sign changes at a defect. The probability distribution, $P(|v_{tet}(i)|)$ (Fig.~S1), also shows the defects all have small volumes compared to those in the helical sections, providing an additional tool for locating and counting defects. To compare this to the random distribution,  $P_R(n)$ (see the Appendix for details), we also calculate a non-equilibrium potential of mean force (PMF),
\begin{equation}
\mathrm{PMF}=-\ln\frac{P(n)}{P_R(n)}\mbox{,}\\
\label{eq:pmf}
\end{equation}
which will be zero if the distributions are the same and negative if helical section lengths are more probable than random. Figure~\ref{fig:helixdist} shows the distributions of odd and even helical sections exhibit similar properties at larger $n$, but at small $n$, the probability of finding an odd-sized helical section rapidly decreases. We also see the presence of sections with $n=1$ that are not predicted by the model, although only with a low probability. At the fastest compression rate, $P(n)$ for the even-sized sections decreases monotonically with increasing $n$, with the exception of the $n=2$ section, but as $d\sigma/dt$ decreases, the distribution develops a shoulder and then becomes bimodal at the slowest compression rate. As a result, the PMF exhibits two minima, one at small $n$ and another at larger $n$, separated by a barrier.


\section{Discussion}
\label{sec:disc}
The inherent structure paradigm applied to hard particle systems describes the statistical mechanics of the fluid by grouping together all the configurations that map the same jammed state into a local basin of attraction. The thermodynamics and dynamics of the fluid and glassy behaviour can then be described in terms of the properties of the accessible basins and how the system moves between them. The goal of the current work is to develop an accurate model of the jammed states for this quasi-one-dimensional model, in terms of helical structures, that improves on the defect crystal model developed previously. In particular, the new model now includes odd-sized helical sections and does not assume a regular arrangement of the defects, which helps account for a broader range of helical arrangements and accurately reproduces the helical geometries found in large packings generated by simulation. The model does miss some possible arrangements, such as the presence of neighbouring defects observed in simulation, but this state appears to be rare and further decreases in probability as the compression rate is decreased.

The model for jammed packings also reveals that the inherent structure landscape has some degree of hierarchical organization, where the basins of configuration space associated with a given arrangement of helical sections split into sub-basins for distinct jammed states. A single helical arrangement represents a basin on the inherent structure landscape because, at high density, there is a significant barrier to moving a particle from one section of helix, which is twisting in one sense, to the next helical section, which is twisting in the opposite sense. For example, particles 3 and 4 in Fig.~\ref{fig:geodef} would need to swap places in terms of the anticlockwise rotation so that the defect becomes located between particles 3 and 4, and particle 3 moves from one helical section to another.  Figure~\ref{fig:oddeven2} then shows a basin can split further into smaller sub-basins, each associated with distinct jammed states that are separated by a small barrier. As the number of helical sections increases, we see an increasingly complex set of minima located within the basin of a given helical arrangement. 

A highly successful mean field theory~\cite{10.1088/1742-5468/2012/10/p10012} of the bulk hard sphere glass transition in high dimensions also predicts the presence of a Gardner transition~\cite{10.1021/jp402235d} between stable and marginally stable glasses that is associated with the repeated fracturing of inherent structure basins into sub-basins at high densities. There is growing evidence to suggest the Gardner transition persists down to three dimensions~\cite{10.1038/ncomms4725,10.1063/1.5097175}, making it relevant to real glassy materials. However, it has also been argued that the physics observed in these lower dimensions results from an avoided transition~\cite{10.1103/PhysRevLett.120.225501}. The quasi-one-dimensional nature of our system suggests that it is unlikely to exhibit a Gardner transition and the types of particle rearrangements involved in the transition between sub-basins are globally concerted motions rather than the highly localized motion thought to be important in bulk glasses. For example, Fig.~\ref{fig:oddeven2} shows that the angle characterizing the structure changes in both helical sections, even though the change in the longer helical section is small and within the  $\{5,5,5,5\}$ helical arrangement, the transition from the jammed packing, $\{\alpha^j_1\}=\{2.2943,2.2943,2.2943,2.2943\}\rightarrow\{2.8104,2.8104,2.8104,2.8104\}$, requires significant changes in all helical sections. However, the properties of the sub-basins still need to be studied in larger systems. The connectivity of the sub-basins also remains to be explored because it is not clear whether the sub-basins show repeated hierarchical organization as expected in bulk glasses. Nevertheless, access to a detailed model of the inherent structure landscape for this system opens up the possibility of understanding how the presence and organization of sub-basins influence the properties of materials at high density.

Our study of jammed packings in small systems shows that $\phi_J$ depends on more than the defect fraction. In particular, the presence of small odd-sized packings tends to lower the packing density. Figure~\ref{fig:njdist} shows that $\phi_J$ varies over a large range for any given $\theta$ and that increasing the fraction of odd sections leads to lower densities, with the least dense packing consisting of alternating sections of two and three particles. Similarly, Fig.~\ref{fig:ddinteration} shows that small, odd-sized helical sections lead to a decrease in $\phi_J$ relative to similar sized, even-sized sections. This inefficiency of packing for small, odd-sized sections has a significant effect on the distribution of helical sections lengths of the large system packing formed in simulation. The LS molecular dynamics scheme continually expands the particles during the trajectory, allowing the system to relax during compression and move to denser basins on the inherent structure landscape that have greater vibrational entropy until the fluid eventually falls out of equilibrium and finally becomes jammed. Figure~\ref{fig:helixdist} shows that the smaller, odd-sized, helical sections are preferentially eliminated from $P(n)$, leading to a non-monotonic distribution with a maximum that moves to larger $n$ as the compression rate is slowed, allowing the system more time to find increasingly dense basins. 

Earlier~\cite{10.1103/physrevlett.115.025702}, it was shown, using a defect crystal model, that this system exhibits a long range defect-defect attraction that results from an increase in $\phi_J$, and hence the vibrational entropy, as two defects approach. Figure~\ref{fig:ddinteration} shows that the same picture persists for a single pair of defects with the improved model, although the inclusion of the odd-sized helical sections complicates the details by adding an oscillation to the attraction. Furthermore, Fig.~\ref{fig:helixdist} clearly shows the PMF for the even-sized sections develops two minima at slow compression rates, one at small $n$ and a second one at larger separations, suggesting the defect-defect attraction leads to a degree of defect pairing. One-dimensional and quasi-one-dimensional systems with particles interacting via short--ranged potentials cannot exhibit a phase transition because there is an entropic advantage in the thermodynamic limit associated with introducing defects that break up long--range order~\cite{10.1016/0031-8914(50)90072-3,lieb,landau}. However, if the interactions between defects become sufficiently long--ranged, phase transitions can occur~\cite{10.1103/PhysRev.187.732,10.1142/S0217979216300188}. Hu et al.~\cite{10.1080/00268976.2018.1479543} recently obtained numerically exact results using the transfer matrix method that shows a transition does not occur in this system, suggesting the configurational entropy, which is related to the number of defects, still dominates at high density, despite the defect-defect attraction driven by the vibrational entropy. The fact that the PMF for the odd-sized helical sections only exhibits a single minimum at large $n$, and that the PMF for the even-sized sections has a large barrier at $n=2$, reducing defect-defect annihilation, may work to counter the attractive interaction and help prevent a transition.  Finally, we also note that $P(n)$, and hence the PMF, are obtained from a non-equilibrium compression of the fluid. It is likely that the jammed states retain some properties of the equilibrium fluid when they fall out of equilibrium, but kinetic effects, such as defect creation and annihilation mechanisms and the ability of defects to diffuse through the structure, could play a role in determining the properties of the packings.

\section{Conclusions}
\label{sec:conc}
A model for the jammed packings of hard spheres confined to a narrow cylindrical channel, allowing only next--nearest contacts, has been developed. It assumes all particles satisfy local packing constraints and describes the structure in terms of helical sections, with alternating twist direction, separated by defects. While it is possible that additional local packing environments could be found, the model provides an improved description of the defects in the packings, as well as packings with odd-sized helical sections, compared to the defect crystal model developed earlier, and accurately reproduces the geometries of jammed packings created by molecular dynamics simulation. The model also confirms the possibility of an entropically driven defect-defect attraction that appears to influence the structure of the particle packings.

\begin{acknowledgments} 
We would like to acknowledge NSERC grants RGPIN--2019--03970 (R. K. B) and RGPIN-2020-04467 (R. J. S.) for financial support. Computational resources were provided by the ICT at the University of Saskatchewan, WestGrid, and Compute Canada.
\end{acknowledgments} 

\appendix
\section{Random Distribution of Helical Section Lengths} 
Here, we derive the random distribution of helical section lengths to obtain an expression for the probability, $P_R(n)$, of finding a helical section of length $n$ in a system of $N$ spheres containing $k$ defects. It is more convenient to develop the analysis in terms of the clusters of the ``bonds" connecting the neighbouring spheres in contact, so that a cluster of $l$ bonds represents a helical section of $n=l+1$ spheres. If $n_0$ is the total number of non-defect bonds and $q_l$ is the number of clusters of with $l$ bonds then we have the following constraints:
\begin{equation}
n_0=\sum_{l=1}^{\infty}lq_l=N-k\mbox{,}\\
\label{eq:con1}
\end{equation}
and
\begin{equation}
k=\sum_{l=1}^{\infty}q_l\mbox{,}\\
\label{eq:con2}
\end{equation}
where the conservation condition,  $N=n_0+k$, was used in the right hand side of Eq.~\ref{eq:con1} and Eq.~\ref{eq:con2} arises from the requirements that each cluster be separated by a defect and defects cannot be adjacent to each other.

The random equilibrium distribution of clusters is obtained by finding the cluster distribution $\{q_l\}$ that maximizes the entropy $S/k_B=\ln\Omega$, subject to the constraints, Eq.~\ref{eq:con1} and Eq.~\ref{eq:con2}, where
\begin{equation}
\Omega=\sum_{\{q_l\}}g\mbox{,}\\
\label{eq:omega}
\end{equation}
and $g$ is the number of distinct ways to distribute the clusters. To obtain an expression for $g$, it is useful to begin by treating all clusters as distinguishable and then remove the over counting of any indistinguishable clusters~\cite{10.1103/physreve.71.026127}. In a system with periodic boundaries, such that particle $N$ contacts particle 1, there are $N$ places to put the first cluster and $(k-1)!$ ways to place the remaining $k-1$ distinguishable clusters. The distinguishability is then removed by dividing by $q_l!$ for each $l$, yielding,
\begin{equation}
g=\frac{N(k-1)!}{\prod_{l=1}^{\infty}q_l!}\mbox{.}\\
\label{eq:g}
\end{equation}

\begin{figure}[]
\includegraphics[]{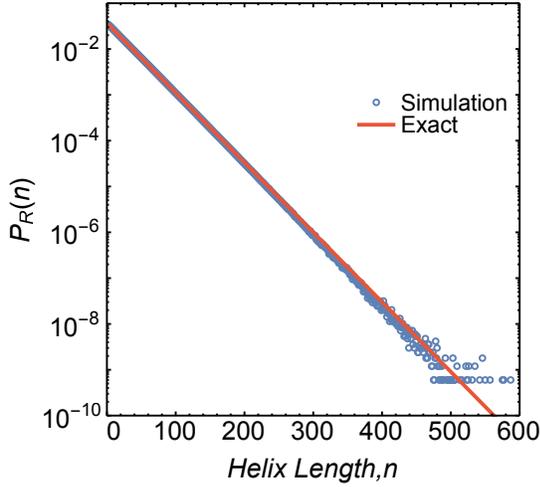}
\caption{Probability, $P_R(n)$, of finding a helical section containing $n$ spheres in a random distribution of defects with $\theta=0.0333$, from simulation (open circles) and exact theory (solid line) given by Eq.~\ref{eq:prn}. }
\label{fig:randomdist}
\end{figure}

The maximum can be obtained by introducing the underdetermined multipliers, $\lambda_1$ and $\lambda_2$, for the two constraints and setting the derivative with respect to $q_m$ to zero, for each $m$,
\begin{multline}
\frac{\partial}{\partial q_m}\left[\ln[\{q_l\}]+\lambda_1\left(N-k-\sum_{l=1}^{\infty}lq_l\right)\right.\\
\left. +\lambda_2\left(k-\sum_{l=1}^{\infty}q_l\right)\right]=0\mbox{.}
\end{multline}
The only terms to survive are those where $m=l$, which gives,
\begin{equation}
q_l=e^{-\lambda_1 l}e^{-\lambda_2}=S_1^lS_2\mbox{,}\\
\label{eq:ql1}
\end{equation}
then using Eq.~\ref{eq:ql1} in the constraints, Eq.~\ref{eq:con1} and Eq.~\ref{eq:con2}, and solving for the unknowns yields,
\begin{equation}
S_1=\frac{N-2k}{N-k}\mbox{,}
\label{eq:s1}
\end{equation}
and
\begin{equation}
S_2=\frac{k^2}{N-2k}\mbox{.}\\
\label{eq:s2}
\end{equation}
Using Eqs~\ref{eq:s1} and \ref{eq:s2} in Eq.~\ref{eq:ql1} and introducing the fraction of defects, $\theta=k/N$, then gives
\begin{equation}
\frac{q_l}{N}=\frac{\theta^2}{1-2\theta}\left[\frac{1-2\theta}{1-\theta}\right]^l\mbox{.}\\
\label{eq:qln}
\end{equation}
Finally, to be consistent with our analysis in the main text, we normalize $P_R(n)$ with respect to the number of defects rather than $N$ and note $n=l+1$, which gives,
\begin{equation}
P_R(n)=\frac{1}{\theta}\frac{q_{n-1}}{N}=\frac{\theta}{1-2\theta}\left[\frac{1-2\theta}{1-\theta}\right]^{n-1}\mbox{.}\\
\label{eq:prn}
\end{equation}

Figure~\ref{fig:randomdist} compares the exact result for $P_R(n)$ (Eq.~\ref{eq:prn}), which is valid in the thermodynamic limit ($N\rightarrow\infty$), with the results obtained from simulations for a system with $N=10000$, where $5\times10^6$ configurations were generated with randomly placed defect bonds such that there were no neighbouring defects bonds and $\theta=0.0333$. The finite--sized system accurately reproduces the exact results for the small helical sections but begins to systematically underestimate $P_R(n)$ for larger sections.


\begin{thebibliography}{62}%
\makeatletter
\providecommand \@ifxundefined [1]{%
 \@ifx{#1\undefined}
}%
\providecommand \@ifnum [1]{%
 \ifnum #1\expandafter \@firstoftwo
 \else \expandafter \@secondoftwo
 \fi
}%
\providecommand \@ifx [1]{%
 \ifx #1\expandafter \@firstoftwo
 \else \expandafter \@secondoftwo
 \fi
}%
\providecommand \natexlab [1]{#1}%
\providecommand \enquote  [1]{``#1''}%
\providecommand \bibnamefont  [1]{#1}%
\providecommand \bibfnamefont [1]{#1}%
\providecommand \citenamefont [1]{#1}%
\providecommand \href@noop [0]{\@secondoftwo}%
\providecommand \href [0]{\begingroup \@sanitize@url \@href}%
\providecommand \@href[1]{\@@startlink{#1}\@@href}%
\providecommand \@@href[1]{\endgroup#1\@@endlink}%
\providecommand \@sanitize@url [0]{\catcode `\\12\catcode `\$12\catcode
  `\&12\catcode `\#12\catcode `\^12\catcode `\_12\catcode `\%12\relax}%
\providecommand \@@startlink[1]{}%
\providecommand \@@endlink[0]{}%
\providecommand \url  [0]{\begingroup\@sanitize@url \@url }%
\providecommand \@url [1]{\endgroup\@href {#1}{\urlprefix }}%
\providecommand \urlprefix  [0]{URL }%
\providecommand \Eprint [0]{\href }%
\providecommand \doibase [0]{https://doi.org/}%
\providecommand \selectlanguage [0]{\@gobble}%
\providecommand \bibinfo  [0]{\@secondoftwo}%
\providecommand \bibfield  [0]{\@secondoftwo}%
\providecommand \translation [1]{[#1]}%
\providecommand \BibitemOpen [0]{}%
\providecommand \bibitemStop [0]{}%
\providecommand \bibitemNoStop [0]{.\EOS\space}%
\providecommand \EOS [0]{\spacefactor3000\relax}%
\providecommand \BibitemShut  [1]{\csname bibitem#1\endcsname}%
\let\auto@bib@innerbib\@empty
\bibitem [{\citenamefont {Kummerfeld}\ \emph {et~al.}(2008)\citenamefont
  {Kummerfeld}, \citenamefont {Hudson},\ and\ \citenamefont
  {Harrowell}}]{10.1021/jp804953r}%
  \BibitemOpen
  \bibfield  {author} {\bibinfo {author} {\bibfnamefont {J.~K.}\ \bibnamefont
  {Kummerfeld}}, \bibinfo {author} {\bibfnamefont {T.~S.}\ \bibnamefont
  {Hudson}},\ and\ \bibinfo {author} {\bibfnamefont {P.}~\bibnamefont
  {Harrowell}},\ }\bibfield  {title} {\bibinfo {title} {{The Densest Packing of
  AB Binary Hard-Sphere Homogeneous Compounds across all Size Ratios}},\ }\href
  {https://doi.org/10.1021/jp804953r} {\bibfield  {journal} {\bibinfo
  {journal} {J. Phys. Chem. B}\ }\textbf {\bibinfo {volume} {112}},\ \bibinfo
  {pages} {10773} (\bibinfo {year} {2008})}\BibitemShut {NoStop}%
\bibitem [{\citenamefont {Hopkins}\ \emph {et~al.}(2012)\citenamefont
  {Hopkins}, \citenamefont {Stillinger},\ and\ \citenamefont
  {Torquato}}]{10.1103/physreve.85.021130}%
  \BibitemOpen
  \bibfield  {author} {\bibinfo {author} {\bibfnamefont {A.~B.}\ \bibnamefont
  {Hopkins}}, \bibinfo {author} {\bibfnamefont {F.~H.}\ \bibnamefont
  {Stillinger}},\ and\ \bibinfo {author} {\bibfnamefont {S.}~\bibnamefont
  {Torquato}},\ }\bibfield  {title} {\bibinfo {title} {{Densest binary sphere
  packings}},\ }\href {https://doi.org/10.1103/physreve.85.021130} {\bibfield
  {journal} {\bibinfo  {journal} {Phys. Rev. E}\ }\textbf {\bibinfo {volume}
  {85}},\ \bibinfo {pages} {021130} (\bibinfo {year} {2012})}\BibitemShut
  {NoStop}%
\bibitem [{\citenamefont {Filion}\ and\ \citenamefont
  {Dijkstra}(2009)}]{10.1103/physreve.79.046714}%
  \BibitemOpen
  \bibfield  {author} {\bibinfo {author} {\bibfnamefont {L.}~\bibnamefont
  {Filion}}\ and\ \bibinfo {author} {\bibfnamefont {M.}~\bibnamefont
  {Dijkstra}},\ }\bibfield  {title} {\bibinfo {title} {{Prediction of binary
  hard-sphere crystal structures}},\ }\href
  {https://doi.org/10.1103/physreve.79.046714} {\bibfield  {journal} {\bibinfo
  {journal} {Phys. Rev. E}\ }\textbf {\bibinfo {volume} {79}},\ \bibinfo
  {pages} {046714} (\bibinfo {year} {2009})}\BibitemShut {NoStop}%
\bibitem [{\citenamefont {Koshoji}\ \emph {et~al.}(2021)\citenamefont
  {Koshoji}, \citenamefont {Kawamura}, \citenamefont {Fukuda},\ and\
  \citenamefont {Ozaki}}]{10.1103/physreve.103.023307}%
  \BibitemOpen
  \bibfield  {author} {\bibinfo {author} {\bibfnamefont {R.}~\bibnamefont
  {Koshoji}}, \bibinfo {author} {\bibfnamefont {M.}~\bibnamefont {Kawamura}},
  \bibinfo {author} {\bibfnamefont {M.}~\bibnamefont {Fukuda}},\ and\ \bibinfo
  {author} {\bibfnamefont {T.}~\bibnamefont {Ozaki}},\ }\bibfield  {title}
  {\bibinfo {title} {{Diverse densest binary sphere packings and phase
  diagram}},\ }\href {https://doi.org/10.1103/physreve.103.023307} {\bibfield
  {journal} {\bibinfo  {journal} {Phys. Rev. E}\ }\textbf {\bibinfo {volume}
  {103}},\ \bibinfo {pages} {023307} (\bibinfo {year} {2021})}\BibitemShut
  {NoStop}%
\bibitem [{\citenamefont {Bernal}\ and\ \citenamefont
  {Mason}(1960)}]{10.1038/188910a0}%
  \BibitemOpen
  \bibfield  {author} {\bibinfo {author} {\bibfnamefont {J.~D.}\ \bibnamefont
  {Bernal}}\ and\ \bibinfo {author} {\bibfnamefont {J.}~\bibnamefont {Mason}},\
  }\bibfield  {title} {\bibinfo {title} {{Packing of Spheres: Co-ordination of
  Randomly Packed Spheres}},\ }\href {https://doi.org/10.1038/188910a0}
  {\bibfield  {journal} {\bibinfo  {journal} {Nature}\ }\textbf {\bibinfo
  {volume} {188}},\ \bibinfo {pages} {910} (\bibinfo {year}
  {1960})}\BibitemShut {NoStop}%
\bibitem [{\citenamefont {Stillinger}\ \emph {et~al.}(1964)\citenamefont
  {Stillinger}, \citenamefont {DiMarzio},\ and\ \citenamefont
  {Kornegay}}]{10.1063/1.1725362}%
  \BibitemOpen
  \bibfield  {author} {\bibinfo {author} {\bibfnamefont {F.~H.}\ \bibnamefont
  {Stillinger}}, \bibinfo {author} {\bibfnamefont {E.~A.}\ \bibnamefont
  {DiMarzio}},\ and\ \bibinfo {author} {\bibfnamefont {R.~L.}\ \bibnamefont
  {Kornegay}},\ }\bibfield  {title} {\bibinfo {title} {{Systematic Approach to
  Explanation of the Rigid Disk Phase Transition}},\ }\href
  {https://doi.org/10.1063/1.1725362} {\bibfield  {journal} {\bibinfo
  {journal} {J. Chem. Phys.}\ }\textbf {\bibinfo {volume} {40}},\ \bibinfo
  {pages} {1564} (\bibinfo {year} {1964})}\BibitemShut {NoStop}%
\bibitem [{\citenamefont {Speedy}(2009)}]{10.1080/00268979809483148}%
  \BibitemOpen
  \bibfield  {author} {\bibinfo {author} {\bibfnamefont {R.~J.}\ \bibnamefont
  {Speedy}},\ }\bibfield  {title} {\bibinfo {title} {{The hard sphere glass
  transition}},\ }\href {https://doi.org/10.1080/00268979809483148} {\bibfield
  {journal} {\bibinfo  {journal} {Molecular Physics}\ }\textbf {\bibinfo
  {volume} {95}},\ \bibinfo {pages} {169} (\bibinfo {year} {2009})}\BibitemShut
  {NoStop}%
\bibitem [{\citenamefont {Liu}\ and\ \citenamefont
  {Nagel}(1998)}]{10.1038/23819}%
  \BibitemOpen
  \bibfield  {author} {\bibinfo {author} {\bibfnamefont {A.~J.}\ \bibnamefont
  {Liu}}\ and\ \bibinfo {author} {\bibfnamefont {S.~R.}\ \bibnamefont
  {Nagel}},\ }\bibfield  {title} {\bibinfo {title} {{Jamming is not just cool
  any more}},\ }\href {https://doi.org/10.1038/23819} {\bibfield  {journal}
  {\bibinfo  {journal} {Nature}\ }\textbf {\bibinfo {volume} {396}},\ \bibinfo
  {pages} {21} (\bibinfo {year} {1998})}\BibitemShut {NoStop}%
\bibitem [{\citenamefont {Torquato}\ \emph {et~al.}(2000)\citenamefont
  {Torquato}, \citenamefont {Truskett},\ and\ \citenamefont
  {Debenedetti}}]{10.1103/physrevlett.84.2064}%
  \BibitemOpen
  \bibfield  {author} {\bibinfo {author} {\bibfnamefont {S.}~\bibnamefont
  {Torquato}}, \bibinfo {author} {\bibfnamefont {T.~M.}\ \bibnamefont
  {Truskett}},\ and\ \bibinfo {author} {\bibfnamefont {P.~G.}\ \bibnamefont
  {Debenedetti}},\ }\bibfield  {title} {\bibinfo {title} {{Is Random Close
  Packing of Spheres Well Defined?}},\ }\href
  {https://doi.org/10.1103/physrevlett.84.2064} {\bibfield  {journal} {\bibinfo
   {journal} {Phys. Rev. Lett.}\ }\textbf {\bibinfo {volume} {84}},\ \bibinfo
  {pages} {2064} (\bibinfo {year} {2000})}\BibitemShut {NoStop}%
\bibitem [{\citenamefont {Parisi}\ and\ \citenamefont
  {Zamponi}(2010)}]{10.1103/revmodphys.82.789}%
  \BibitemOpen
  \bibfield  {author} {\bibinfo {author} {\bibfnamefont {G.}~\bibnamefont
  {Parisi}}\ and\ \bibinfo {author} {\bibfnamefont {F.}~\bibnamefont
  {Zamponi}},\ }\bibfield  {title} {\bibinfo {title} {{Mean-field theory of
  hard sphere glasses and jamming}},\ }\href
  {https://doi.org/10.1103/revmodphys.82.789} {\bibfield  {journal} {\bibinfo
  {journal} {Rev. Mod. Phys.}\ }\textbf {\bibinfo {volume} {82}},\ \bibinfo
  {pages} {789} (\bibinfo {year} {2010})}\BibitemShut {NoStop}%
\bibitem [{\citenamefont {Torquato}\ and\ \citenamefont
  {Stillinger}(2010)}]{10.1103/revmodphys.82.2633}%
  \BibitemOpen
  \bibfield  {author} {\bibinfo {author} {\bibfnamefont {S.}~\bibnamefont
  {Torquato}}\ and\ \bibinfo {author} {\bibfnamefont {F.~H.}\ \bibnamefont
  {Stillinger}},\ }\bibfield  {title} {\bibinfo {title} {{Jammed hard-particle
  packings: From Kepler to Bernal and beyond}},\ }\href
  {https://doi.org/10.1103/revmodphys.82.2633} {\bibfield  {journal} {\bibinfo
  {journal} {Rev. Mod. Phys.}\ }\textbf {\bibinfo {volume} {82}},\ \bibinfo
  {pages} {2633} (\bibinfo {year} {2010})}\BibitemShut {NoStop}%
\bibitem [{\citenamefont {Torquato}(2018)}]{10.1063/1.5036657}%
  \BibitemOpen
  \bibfield  {author} {\bibinfo {author} {\bibfnamefont {S.}~\bibnamefont
  {Torquato}},\ }\bibfield  {title} {\bibinfo {title} {{Perspective: Basic
  understanding of condensed phases of matter via packing models}},\ }\href
  {https://doi.org/10.1063/1.5036657} {\bibfield  {journal} {\bibinfo
  {journal} {J. Chem. Phys.}\ }\textbf {\bibinfo {volume} {149}},\ \bibinfo
  {pages} {020901} (\bibinfo {year} {2018})}\BibitemShut {NoStop}%
\bibitem [{\citenamefont {Pusey}\ and\ \citenamefont
  {Megen}(1986)}]{10.1038/320340a0}%
  \BibitemOpen
  \bibfield  {author} {\bibinfo {author} {\bibfnamefont {P.~N.}\ \bibnamefont
  {Pusey}}\ and\ \bibinfo {author} {\bibfnamefont {W.~v.}\ \bibnamefont
  {Megen}},\ }\bibfield  {title} {\bibinfo {title} {{Phase behaviour of
  concentrated suspensions of nearly hard colloidal spheres}},\ }\href
  {https://doi.org/10.1038/320340a0} {\bibfield  {journal} {\bibinfo  {journal}
  {Nature}\ }\textbf {\bibinfo {volume} {320}},\ \bibinfo {pages} {340}
  (\bibinfo {year} {1986})}\BibitemShut {NoStop}%
\bibitem [{\citenamefont {Edwards}\ and\ \citenamefont
  {Oakeshott}(1989)}]{10.1016/0378-4371(89)90034-4}%
  \BibitemOpen
  \bibfield  {author} {\bibinfo {author} {\bibfnamefont {S.}~\bibnamefont
  {Edwards}}\ and\ \bibinfo {author} {\bibfnamefont {R.}~\bibnamefont
  {Oakeshott}},\ }\bibfield  {title} {\bibinfo {title} {{Theory of powders}},\
  }\href {https://doi.org/10.1016/0378-4371(89)90034-4} {\bibfield  {journal}
  {\bibinfo  {journal} {Physica A}\ }\textbf {\bibinfo {volume} {157}},\
  \bibinfo {pages} {1080} (\bibinfo {year} {1989})}\BibitemShut {NoStop}%
\bibitem [{\citenamefont {Bi}\ \emph {et~al.}(2015)\citenamefont {Bi},
  \citenamefont {Henkes}, \citenamefont {Daniels},\ and\ \citenamefont
  {Chakraborty}}]{10.1146/annurev-conmatphys-031214-014336}%
  \BibitemOpen
  \bibfield  {author} {\bibinfo {author} {\bibfnamefont {D.}~\bibnamefont
  {Bi}}, \bibinfo {author} {\bibfnamefont {S.}~\bibnamefont {Henkes}}, \bibinfo
  {author} {\bibfnamefont {K.~E.}\ \bibnamefont {Daniels}},\ and\ \bibinfo
  {author} {\bibfnamefont {B.}~\bibnamefont {Chakraborty}},\ }\bibfield
  {title} {\bibinfo {title} {{The Statistical Physics of Athermal Materials}},\
  }\href {https://doi.org/10.1146/annurev-conmatphys-031214-014336} {\bibfield
  {journal} {\bibinfo  {journal} {Ann. Rev. Condens. Matter Phys.}\ }\textbf
  {\bibinfo {volume} {6}},\ \bibinfo {pages} {63} (\bibinfo {year}
  {2015})}\BibitemShut {NoStop}%
\bibitem [{\citenamefont {Mezard}\ and\ \citenamefont
  {Montanari}(2009)}]{mezard_2009}%
  \BibitemOpen
  \bibfield  {author} {\bibinfo {author} {\bibfnamefont {M.}~\bibnamefont
  {Mezard}}\ and\ \bibinfo {author} {\bibfnamefont {A.}~\bibnamefont
  {Montanari}},\ }\bibfield  {title} {\bibinfo {title} {{Information, Physics,
  and Computation}},\ }\href@noop {} {\bibfield  {journal} {\bibinfo  {journal}
  {Oxford University Press, Inc.}\ } (\bibinfo {year} {2009})}\BibitemShut
  {NoStop}%
\bibitem [{\citenamefont {Hales}(2005)}]{10.4007/annals.2005.162.1065}%
  \BibitemOpen
  \bibfield  {author} {\bibinfo {author} {\bibfnamefont {T.}~\bibnamefont
  {Hales}},\ }\bibfield  {title} {\bibinfo {title} {{A proof of the Kepler
  conjecture}},\ }\href {https://doi.org/10.4007/annals.2005.162.1065}
  {\bibfield  {journal} {\bibinfo  {journal} {Ann. Math.}\ }\textbf {\bibinfo
  {volume} {162}},\ \bibinfo {pages} {1065} (\bibinfo {year}
  {2005})}\BibitemShut {NoStop}%
\bibitem [{\citenamefont {Bowles}\ and\ \citenamefont
  {Speedy}(1999)}]{10.1016/s0378-4371(98)00404-x}%
  \BibitemOpen
  \bibfield  {author} {\bibinfo {author} {\bibfnamefont {R.~K.}\ \bibnamefont
  {Bowles}}\ and\ \bibinfo {author} {\bibfnamefont {R.~J.}\ \bibnamefont
  {Speedy}},\ }\bibfield  {title} {\bibinfo {title} {{Five discs in a box}},\
  }\href {https://doi.org/10.1016/s0378-4371(98)00404-x} {\bibfield  {journal}
  {\bibinfo  {journal} {Physica A}\ }\textbf {\bibinfo {volume} {262}},\
  \bibinfo {pages} {76} (\bibinfo {year} {1999})}\BibitemShut {NoStop}%
\bibitem [{\citenamefont {Weeks}\ and\ \citenamefont
  {Criddle}(2020)}]{2001.11635}%
  \BibitemOpen
  \bibfield  {author} {\bibinfo {author} {\bibfnamefont {E.~R.}\ \bibnamefont
  {Weeks}}\ and\ \bibinfo {author} {\bibfnamefont {K.}~\bibnamefont
  {Criddle}},\ }\bibfield  {title} {\bibinfo {title} {{Visualizing free-energy
  landscapes for four hard disks}},\ }\href
  {https://doi.org/10.1103/physreve.102.062153} {\bibfield  {journal} {\bibinfo
   {journal} {Phys. Rev. E}\ }\textbf {\bibinfo {volume} {102}},\ \bibinfo
  {pages} {062153} (\bibinfo {year} {2020})}\BibitemShut {NoStop}%
\bibitem [{\citenamefont {Bowles}\ and\ \citenamefont
  {Saika-Voivod}(2006)}]{10.1103/PhysRevE.73.011503}%
  \BibitemOpen
  \bibfield  {author} {\bibinfo {author} {\bibfnamefont {R.~K.}\ \bibnamefont
  {Bowles}}\ and\ \bibinfo {author} {\bibfnamefont {I.}~\bibnamefont
  {Saika-Voivod}},\ }\bibfield  {title} {\bibinfo {title} {{Landscapes, dynamic
  heterogeneity, and kinetic facilitation in a simple off-lattice model}},\
  }\href {https://doi.org/10.1103/PhysRevE.73.011503} {\bibfield  {journal}
  {\bibinfo  {journal} {Phys. Rev. E}\ }\textbf {\bibinfo {volume} {73}},\
  \bibinfo {pages} {011503} (\bibinfo {year} {2006})}\BibitemShut {NoStop}%
\bibitem [{\citenamefont {Ashwin}\ and\ \citenamefont
  {Bowles}(2009)}]{10.1103/PhysRevLett.102.235701}%
  \BibitemOpen
  \bibfield  {author} {\bibinfo {author} {\bibfnamefont {S.~S.}\ \bibnamefont
  {Ashwin}}\ and\ \bibinfo {author} {\bibfnamefont {R.~K.}\ \bibnamefont
  {Bowles}},\ }\bibfield  {title} {\bibinfo {title} {{Complete Jamming
  Landscape of Confined Hard Discs}},\ }\href
  {https://doi.org/10.1103/PhysRevLett.102.235701} {\bibfield  {journal}
  {\bibinfo  {journal} {Phys. Rev. Lett.}\ }\textbf {\bibinfo {volume} {102}},\
  \bibinfo {pages} {235701} (\bibinfo {year} {2009})}\BibitemShut {NoStop}%
\bibitem [{\citenamefont {Yamchi}\ \emph {et~al.}(2012)\citenamefont {Yamchi},
  \citenamefont {Ashwin},\ and\ \citenamefont
  {Bowles}}]{10.1103/PhysRevLett.109.225701}%
  \BibitemOpen
  \bibfield  {author} {\bibinfo {author} {\bibfnamefont {M.~Z.}\ \bibnamefont
  {Yamchi}}, \bibinfo {author} {\bibfnamefont {S.~S.}\ \bibnamefont {Ashwin}},\
  and\ \bibinfo {author} {\bibfnamefont {R.~K.}\ \bibnamefont {Bowles}},\
  }\bibfield  {title} {\bibinfo {title} {{Fragile-Strong Fluid Crossover and
  Universal Relaxation Times in a Confined Hard-Disk Fluid}},\ }\href
  {https://doi.org/10.1103/physrevlett.109.225701} {\bibfield  {journal}
  {\bibinfo  {journal} {Phys. Rev. Lett.}\ }\textbf {\bibinfo {volume} {109}},\
  \bibinfo {pages} {225701} (\bibinfo {year} {2012})}\BibitemShut {NoStop}%
\bibitem [{\citenamefont {Ashwin}\ \emph {et~al.}(2013)\citenamefont {Ashwin},
  \citenamefont {Yamchi},\ and\ \citenamefont
  {Bowles}}]{10.1103/PhysRevLett.110.145701}%
  \BibitemOpen
  \bibfield  {author} {\bibinfo {author} {\bibfnamefont {S.~S.}\ \bibnamefont
  {Ashwin}}, \bibinfo {author} {\bibfnamefont {M.}~\bibnamefont {Yamchi}},\
  and\ \bibinfo {author} {\bibfnamefont {R.~K.}\ \bibnamefont {Bowles}},\
  }\bibfield  {title} {\bibinfo {title} {{Inherent Structure Landscape
  Connection between Liquids, Granular Materials, and the Jamming Phase
  Diagram}},\ }\href {https://doi.org/10.1103/PhysRevLett.110.145701}
  {\bibfield  {journal} {\bibinfo  {journal} {Phys. Rev. Lett.}\ }\textbf
  {\bibinfo {volume} {110}},\ \bibinfo {pages} {145701} (\bibinfo {year}
  {2013})}\BibitemShut {NoStop}%
\bibitem [{\citenamefont {Godfrey}\ and\ \citenamefont
  {Moore}(2014)}]{10.1103/PhysRevE.89.032111}%
  \BibitemOpen
  \bibfield  {author} {\bibinfo {author} {\bibfnamefont {M.~J.}\ \bibnamefont
  {Godfrey}}\ and\ \bibinfo {author} {\bibfnamefont {M.~A.}\ \bibnamefont
  {Moore}},\ }\bibfield  {title} {\bibinfo {title} {{Static and dynamical
  properties of a hard-disk fluid confined to a narrow channel}},\ }\href
  {https://doi.org/10.1103/PhysRevE.89.032111} {\bibfield  {journal} {\bibinfo
  {journal} {Phys. Rev. E}\ }\textbf {\bibinfo {volume} {89}},\ \bibinfo
  {pages} {032111} (\bibinfo {year} {2014})}\BibitemShut {NoStop}%
\bibitem [{\citenamefont {Godfrey}\ and\ \citenamefont
  {Moore}(2015)}]{10.1103/PhysRevE.91.022120}%
  \BibitemOpen
  \bibfield  {author} {\bibinfo {author} {\bibfnamefont {M.~J.}\ \bibnamefont
  {Godfrey}}\ and\ \bibinfo {author} {\bibfnamefont {M.~A.}\ \bibnamefont
  {Moore}},\ }\bibfield  {title} {\bibinfo {title} {{Understanding the ideal
  glass transition: Lessons from an equilibrium study of hard disks in a
  channel}},\ }\href {https://doi.org/10.1103/physreve.91.022120} {\bibfield
  {journal} {\bibinfo  {journal} {Phys. Rev. E}\ }\textbf {\bibinfo {volume}
  {91}},\ \bibinfo {pages} {022120} (\bibinfo {year} {2015})}\BibitemShut
  {NoStop}%
\bibitem [{\citenamefont {Yamchi}\ \emph {et~al.}(2015)\citenamefont {Yamchi},
  \citenamefont {Ashwin},\ and\ \citenamefont
  {Bowles}}]{10.1103/physreve.91.022301}%
  \BibitemOpen
  \bibfield  {author} {\bibinfo {author} {\bibfnamefont {M.~Z.}\ \bibnamefont
  {Yamchi}}, \bibinfo {author} {\bibfnamefont {S.~S.}\ \bibnamefont {Ashwin}},\
  and\ \bibinfo {author} {\bibfnamefont {R.~K.}\ \bibnamefont {Bowles}},\
  }\bibfield  {title} {\bibinfo {title} {{Inherent structures, fragility, and
  jamming: Insights from quasi-one-dimensional hard disks}},\ }\href
  {https://doi.org/10.1103/physreve.91.022301} {\bibfield  {journal} {\bibinfo
  {journal} {Phys. Rev. E}\ }\textbf {\bibinfo {volume} {91}},\ \bibinfo
  {pages} {022301} (\bibinfo {year} {2015})}\BibitemShut {NoStop}%
\bibitem [{\citenamefont {Robinson}\ \emph {et~al.}(2016)\citenamefont
  {Robinson}, \citenamefont {Godfrey},\ and\ \citenamefont
  {Moore}}]{10.1103/PhysRevE.93.032101}%
  \BibitemOpen
  \bibfield  {author} {\bibinfo {author} {\bibfnamefont {J.~F.}\ \bibnamefont
  {Robinson}}, \bibinfo {author} {\bibfnamefont {M.~J.}\ \bibnamefont
  {Godfrey}},\ and\ \bibinfo {author} {\bibfnamefont {M.~A.}\ \bibnamefont
  {Moore}},\ }\bibfield  {title} {\bibinfo {title} {{Glasslike behavior of a
  hard-disk fluid confined to a narrow channel}},\ }\href
  {https://doi.org/10.1103/physreve.93.032101} {\bibfield  {journal} {\bibinfo
  {journal} {Phys. Rev. E}\ }\textbf {\bibinfo {volume} {93}},\ \bibinfo
  {pages} {032101} (\bibinfo {year} {2016})}\BibitemShut {NoStop}%
\bibitem [{\citenamefont {Godfrey}\ and\ \citenamefont
  {Moore}(2018)}]{10.1103/PhysRevLett.121.075503}%
  \BibitemOpen
  \bibfield  {author} {\bibinfo {author} {\bibfnamefont {M.}~\bibnamefont
  {Godfrey}}\ and\ \bibinfo {author} {\bibfnamefont {M.}~\bibnamefont
  {Moore}},\ }\bibfield  {title} {\bibinfo {title} {{Absence of Hyperuniformity
  in Amorphous Hard-Sphere Packings of Nonvanishing Complexity}},\ }\href
  {https://doi.org/10.1103/PhysRevLett.121.075503} {\bibfield  {journal}
  {\bibinfo  {journal} {Phys. Rev. Lett.}\ }\textbf {\bibinfo {volume} {121}},\
  \bibinfo {pages} {075503} (\bibinfo {year} {2018})}\BibitemShut {NoStop}%
\bibitem [{\citenamefont {Bowles}\ and\ \citenamefont
  {Ashwin}(2011)}]{10.1103/PhysRevE.83.031302}%
  \BibitemOpen
  \bibfield  {author} {\bibinfo {author} {\bibfnamefont {R.~K.}\ \bibnamefont
  {Bowles}}\ and\ \bibinfo {author} {\bibfnamefont {S.~S.}\ \bibnamefont
  {Ashwin}},\ }\bibfield  {title} {\bibinfo {title} {{Edwards entropy and
  compactivity in a model of granular matter}},\ }\href
  {https://doi.org/10.1103/PhysRevE.83.031302} {\bibfield  {journal} {\bibinfo
  {journal} {Phys. Rev. E}\ }\textbf {\bibinfo {volume} {83}},\ \bibinfo
  {pages} {031302} (\bibinfo {year} {2011})}\BibitemShut {NoStop}%
\bibitem [{\citenamefont {Hicks}\ \emph {et~al.}(2018)\citenamefont {Hicks},
  \citenamefont {Wheatley}, \citenamefont {Godfrey},\ and\ \citenamefont
  {Moore}}]{10.1103/PhysRevLett.120.225501}%
  \BibitemOpen
  \bibfield  {author} {\bibinfo {author} {\bibfnamefont {C.}~\bibnamefont
  {Hicks}}, \bibinfo {author} {\bibfnamefont {M.}~\bibnamefont {Wheatley}},
  \bibinfo {author} {\bibfnamefont {M.}~\bibnamefont {Godfrey}},\ and\ \bibinfo
  {author} {\bibfnamefont {M.}~\bibnamefont {Moore}},\ }\bibfield  {title}
  {\bibinfo {title} {{Gardner Transition in Physical Dimensions}},\ }\href
  {https://doi.org/10.1103/PhysRevLett.120.225501} {\bibfield  {journal}
  {\bibinfo  {journal} {Phys. Rev. Lett.}\ }\textbf {\bibinfo {volume} {120}},\
  \bibinfo {pages} {225501} (\bibinfo {year} {2018})}\BibitemShut {NoStop}%
\bibitem [{\citenamefont {Zhang}\ \emph {et~al.}(2020)\citenamefont {Zhang},
  \citenamefont {Godfrey},\ and\ \citenamefont
  {Moore}}]{10.1103/physreve.102.042614}%
  \BibitemOpen
  \bibfield  {author} {\bibinfo {author} {\bibfnamefont {Y.}~\bibnamefont
  {Zhang}}, \bibinfo {author} {\bibfnamefont {M.~J.}\ \bibnamefont {Godfrey}},\
  and\ \bibinfo {author} {\bibfnamefont {M.~A.}\ \bibnamefont {Moore}},\
  }\bibfield  {title} {\bibinfo {title} {{Marginally jammed states of hard
  disks in a one-dimensional channel}},\ }\href
  {https://doi.org/10.1103/physreve.102.042614} {\bibfield  {journal} {\bibinfo
   {journal} {Phys. Rev. E}\ }\textbf {\bibinfo {volume} {102}},\ \bibinfo
  {pages} {042614} (\bibinfo {year} {2020})}\BibitemShut {NoStop}%
\bibitem [{\citenamefont {Mughal}\ \emph {et~al.}(2011)\citenamefont {Mughal},
  \citenamefont {Chan},\ and\ \citenamefont
  {Weaire}}]{10.1103/physrevlett.106.115704}%
  \BibitemOpen
  \bibfield  {author} {\bibinfo {author} {\bibfnamefont {A.}~\bibnamefont
  {Mughal}}, \bibinfo {author} {\bibfnamefont {H.~K.}\ \bibnamefont {Chan}},\
  and\ \bibinfo {author} {\bibfnamefont {D.}~\bibnamefont {Weaire}},\
  }\bibfield  {title} {\bibinfo {title} {{Phyllotactic Description of Hard
  Sphere Packing in Cylindrical Channels}},\ }\href
  {https://doi.org/10.1103/physrevlett.106.115704} {\bibfield  {journal}
  {\bibinfo  {journal} {Phys. Rev. Lett.}\ }\textbf {\bibinfo {volume} {106}},\
  \bibinfo {pages} {115704} (\bibinfo {year} {2011})}\BibitemShut {NoStop}%
\bibitem [{\citenamefont {Mughal}\ and\ \citenamefont
  {Weaire}(2014)}]{10.1103/physreve.89.042307}%
  \BibitemOpen
  \bibfield  {author} {\bibinfo {author} {\bibfnamefont {A.}~\bibnamefont
  {Mughal}}\ and\ \bibinfo {author} {\bibfnamefont {D.}~\bibnamefont
  {Weaire}},\ }\bibfield  {title} {\bibinfo {title} {{Theory of cylindrical
  dense packings of disks}},\ }\href
  {https://doi.org/10.1103/physreve.89.042307} {\bibfield  {journal} {\bibinfo
  {journal} {Phys. Rev. E}\ }\textbf {\bibinfo {volume} {89}},\ \bibinfo
  {pages} {042307} (\bibinfo {year} {2014})}\BibitemShut {NoStop}%
\bibitem [{\citenamefont {Chan}\ \emph {et~al.}(2019)\citenamefont {Chan},
  \citenamefont {Wang},\ and\ \citenamefont {Han}}]{10.1063/1.5131318}%
  \BibitemOpen
  \bibfield  {author} {\bibinfo {author} {\bibfnamefont {H.-K.}\ \bibnamefont
  {Chan}}, \bibinfo {author} {\bibfnamefont {Y.}~\bibnamefont {Wang}},\ and\
  \bibinfo {author} {\bibfnamefont {H.}~\bibnamefont {Han}},\ }\bibfield
  {title} {\bibinfo {title} {{Densest helical structures of hard spheres in
  narrow confinement: An analytic derivation}},\ }\href
  {https://doi.org/10.1063/1.5131318} {\bibfield  {journal} {\bibinfo
  {journal} {AIP Advances}\ }\textbf {\bibinfo {volume} {9}},\ \bibinfo {pages}
  {125118} (\bibinfo {year} {2019})}\BibitemShut {NoStop}%
\bibitem [{\citenamefont {Gordillo}\ \emph {et~al.}(2006)\citenamefont
  {Gordillo}, \citenamefont {Martínez-Haya},\ and\ \citenamefont
  {Romero-Enrique}}]{10.1063/1.2358135}%
  \BibitemOpen
  \bibfield  {author} {\bibinfo {author} {\bibfnamefont {M.~C.}\ \bibnamefont
  {Gordillo}}, \bibinfo {author} {\bibfnamefont {B.}~\bibnamefont
  {Martínez-Haya}},\ and\ \bibinfo {author} {\bibfnamefont {J.~M.}\
  \bibnamefont {Romero-Enrique}},\ }\bibfield  {title} {\bibinfo {title}
  {{Freezing of hard spheres confined in narrow cylindrical pores}},\ }\href
  {https://doi.org/10.1063/1.2358135} {\bibfield  {journal} {\bibinfo
  {journal} {J. Chem. Phys.}\ }\textbf {\bibinfo {volume} {125}},\ \bibinfo
  {pages} {144702} (\bibinfo {year} {2006})}\BibitemShut {NoStop}%
\bibitem [{\citenamefont {Mughal}\ \emph {et~al.}(2012)\citenamefont {Mughal},
  \citenamefont {Chan}, \citenamefont {Weaire},\ and\ \citenamefont
  {Hutzler}}]{10.1103/physreve.85.051305}%
  \BibitemOpen
  \bibfield  {author} {\bibinfo {author} {\bibfnamefont {A.}~\bibnamefont
  {Mughal}}, \bibinfo {author} {\bibfnamefont {H.~K.}\ \bibnamefont {Chan}},
  \bibinfo {author} {\bibfnamefont {D.}~\bibnamefont {Weaire}},\ and\ \bibinfo
  {author} {\bibfnamefont {S.}~\bibnamefont {Hutzler}},\ }\bibfield  {title}
  {\bibinfo {title} {{Dense packings of spheres in cylinders: Simulations}},\
  }\href {https://doi.org/10.1103/physreve.85.051305} {\bibfield  {journal}
  {\bibinfo  {journal} {Phys. Rev. E}\ }\textbf {\bibinfo {volume} {85}},\
  \bibinfo {pages} {051305} (\bibinfo {year} {2012})}\BibitemShut {NoStop}%
\bibitem [{\citenamefont {Durán-Olivencia}\ and\ \citenamefont
  {Gordillo}(2009)}]{10.1103/physreve.79.061111}%
  \BibitemOpen
  \bibfield  {author} {\bibinfo {author} {\bibfnamefont {F.~J.}\ \bibnamefont
  {Durán-Olivencia}}\ and\ \bibinfo {author} {\bibfnamefont {M.~C.}\
  \bibnamefont {Gordillo}},\ }\bibfield  {title} {\bibinfo {title} {{Ordering
  of hard spheres inside hard cylindrical pores}},\ }\href
  {https://doi.org/10.1103/physreve.79.061111} {\bibfield  {journal} {\bibinfo
  {journal} {Phys. Rev. E}\ }\textbf {\bibinfo {volume} {79}},\ \bibinfo
  {pages} {061111} (\bibinfo {year} {2009})}\BibitemShut {NoStop}%
\bibitem [{\citenamefont {Fu}\ \emph {et~al.}(2016)\citenamefont {Fu},
  \citenamefont {Steinhardt}, \citenamefont {Zhao}, \citenamefont {Socolar},\
  and\ \citenamefont {Charbonneau}}]{10.1039/C5SM02875B}%
  \BibitemOpen
  \bibfield  {author} {\bibinfo {author} {\bibfnamefont {L.}~\bibnamefont
  {Fu}}, \bibinfo {author} {\bibfnamefont {W.}~\bibnamefont {Steinhardt}},
  \bibinfo {author} {\bibfnamefont {H.}~\bibnamefont {Zhao}}, \bibinfo {author}
  {\bibfnamefont {J.~E.}\ \bibnamefont {Socolar}},\ and\ \bibinfo {author}
  {\bibfnamefont {P.}~\bibnamefont {Charbonneau}},\ }\bibfield  {title}
  {\bibinfo {title} {{Hard sphere packings within cylinders}},\ }\href
  {https://doi.org/10.1039/C5SM02875B} {\bibfield  {journal} {\bibinfo
  {journal} {Soft Matter}\ }\textbf {\bibinfo {volume} {12}},\ \bibinfo {pages}
  {2505} (\bibinfo {year} {2016})}\BibitemShut {NoStop}%
\bibitem [{\citenamefont {Erickson}(1973)}]{10.1126/science.181.4101.705}%
  \BibitemOpen
  \bibfield  {author} {\bibinfo {author} {\bibfnamefont {R.~O.}\ \bibnamefont
  {Erickson}},\ }\bibfield  {title} {\bibinfo {title} {{Tubular Packing of
  Spheres in Biological Fine Structure}},\ }\href
  {https://doi.org/10.1126/science.181.4101.705} {\bibfield  {journal}
  {\bibinfo  {journal} {Science}\ }\textbf {\bibinfo {volume} {181}},\ \bibinfo
  {pages} {705} (\bibinfo {year} {1973})}\BibitemShut {NoStop}%
\bibitem [{\citenamefont {Harris}\ and\ \citenamefont
  {Erickson}(1980)}]{10.1016/0022-5193(80)90290-8}%
  \BibitemOpen
  \bibfield  {author} {\bibinfo {author} {\bibfnamefont {W.~F.}\ \bibnamefont
  {Harris}}\ and\ \bibinfo {author} {\bibfnamefont {R.~O.}\ \bibnamefont
  {Erickson}},\ }\bibfield  {title} {\bibinfo {title} {{Tubular arrays of
  spheres: Geometry, continuous and discontinuous contraction, and the role of
  moving dislocations}},\ }\href {https://doi.org/10.1016/0022-5193(80)90290-8}
  {\bibfield  {journal} {\bibinfo  {journal} {J. Theor. Biol.}\ }\textbf
  {\bibinfo {volume} {83}},\ \bibinfo {pages} {215} (\bibinfo {year}
  {1980})}\BibitemShut {NoStop}%
\bibitem [{\citenamefont {Pickett}\ \emph {et~al.}(2000)\citenamefont
  {Pickett}, \citenamefont {Gross},\ and\ \citenamefont
  {Okuyama}}]{10.1103/physrevlett.85.3652}%
  \BibitemOpen
  \bibfield  {author} {\bibinfo {author} {\bibfnamefont {G.~T.}\ \bibnamefont
  {Pickett}}, \bibinfo {author} {\bibfnamefont {M.}~\bibnamefont {Gross}},\
  and\ \bibinfo {author} {\bibfnamefont {H.}~\bibnamefont {Okuyama}},\
  }\bibfield  {title} {\bibinfo {title} {{Spontaneous Chirality in Simple
  Systems}},\ }\href {https://doi.org/10.1103/physrevlett.85.3652} {\bibfield
  {journal} {\bibinfo  {journal} {Phys. Rev. Lett.}\ }\textbf {\bibinfo
  {volume} {85}},\ \bibinfo {pages} {3652} (\bibinfo {year}
  {2000})}\BibitemShut {NoStop}%
\bibitem [{\citenamefont {Mickelson}\ \emph {et~al.}(2003)\citenamefont
  {Mickelson}, \citenamefont {Aloni}, \citenamefont {Han}, \citenamefont
  {Cumings},\ and\ \citenamefont {Zettl}}]{10.1126/science.1082346}%
  \BibitemOpen
  \bibfield  {author} {\bibinfo {author} {\bibfnamefont {W.}~\bibnamefont
  {Mickelson}}, \bibinfo {author} {\bibfnamefont {S.}~\bibnamefont {Aloni}},
  \bibinfo {author} {\bibfnamefont {W.-Q.}\ \bibnamefont {Han}}, \bibinfo
  {author} {\bibfnamefont {J.}~\bibnamefont {Cumings}},\ and\ \bibinfo {author}
  {\bibfnamefont {A.}~\bibnamefont {Zettl}},\ }\bibfield  {title} {\bibinfo
  {title} {{Packing C60 in Boron Nitride Nanotubes}},\ }\href
  {https://doi.org/10.1126/science.1082346} {\bibfield  {journal} {\bibinfo
  {journal} {Science}\ }\textbf {\bibinfo {volume} {300}},\ \bibinfo {pages}
  {467} (\bibinfo {year} {2003})}\BibitemShut {NoStop}%
\bibitem [{\citenamefont {Khlobystov}\ \emph {et~al.}(2004)\citenamefont
  {Khlobystov}, \citenamefont {Britz}, \citenamefont {Ardavan},\ and\
  \citenamefont {Briggs}}]{10.1103/PhysRevLett.92.245507}%
  \BibitemOpen
  \bibfield  {author} {\bibinfo {author} {\bibfnamefont {A.~N.}\ \bibnamefont
  {Khlobystov}}, \bibinfo {author} {\bibfnamefont {D.~A.}\ \bibnamefont
  {Britz}}, \bibinfo {author} {\bibfnamefont {A.}~\bibnamefont {Ardavan}},\
  and\ \bibinfo {author} {\bibfnamefont {G.~A.~D.}\ \bibnamefont {Briggs}},\
  }\bibfield  {title} {\bibinfo {title} {{Observation of Ordered Phases of
  Fullerenes in Carbon Nanotubes}},\ }\href
  {https://doi.org/10.1103/physrevlett.92.245507} {\bibfield  {journal}
  {\bibinfo  {journal} {Phys. Rev. Lett.}\ }\textbf {\bibinfo {volume} {92}},\
  \bibinfo {pages} {245507} (\bibinfo {year} {2004})}\BibitemShut {NoStop}%
\bibitem [{\citenamefont {Jiang}\ \emph {et~al.}(2013)\citenamefont {Jiang},
  \citenamefont {de~Folter}, \citenamefont {Huang}, \citenamefont {Philipse},
  \citenamefont {Kegel},\ and\ \citenamefont
  {Petukov}}]{10.1002/anie.201209767}%
  \BibitemOpen
  \bibfield  {author} {\bibinfo {author} {\bibfnamefont {L.}~\bibnamefont
  {Jiang}}, \bibinfo {author} {\bibfnamefont {J.~W.~J.}\ \bibnamefont
  {de~Folter}}, \bibinfo {author} {\bibfnamefont {J.}~\bibnamefont {Huang}},
  \bibinfo {author} {\bibfnamefont {A.~P.}\ \bibnamefont {Philipse}}, \bibinfo
  {author} {\bibfnamefont {W.~K.}\ \bibnamefont {Kegel}},\ and\ \bibinfo
  {author} {\bibfnamefont {A.~V.}\ \bibnamefont {Petukov}},\ }\bibfield
  {title} {\bibinfo {title} {{Helical Colloidal Sphere Structures through
  Thermo‐Reversible Co‐Assembly with Molecular Microtubes }},\ }\href
  {https://doi.org/10.1002/anie.201209767} {\bibfield  {journal} {\bibinfo
  {journal} {Angew. Chem. Int. Ed.}\ }\textbf {\bibinfo {volume} {52}},\
  \bibinfo {pages} {3364} (\bibinfo {year} {2013})}\BibitemShut {NoStop}%
\bibitem [{\citenamefont {Fu}\ \emph {et~al.}(2017)\citenamefont {Fu},
  \citenamefont {Bian}, \citenamefont {Shields}, \citenamefont {Cruz},
  \citenamefont {López},\ and\ \citenamefont
  {Charbonneau}}]{10.1039/C7SM00316A}%
  \BibitemOpen
  \bibfield  {author} {\bibinfo {author} {\bibfnamefont {L.}~\bibnamefont
  {Fu}}, \bibinfo {author} {\bibfnamefont {C.}~\bibnamefont {Bian}}, \bibinfo
  {author} {\bibfnamefont {W.~C.}\ \bibnamefont {Shields}}, \bibinfo {author}
  {\bibfnamefont {D.~F.}\ \bibnamefont {Cruz}}, \bibinfo {author}
  {\bibfnamefont {G.~P.}\ \bibnamefont {López}},\ and\ \bibinfo {author}
  {\bibfnamefont {P.}~\bibnamefont {Charbonneau}},\ }\bibfield  {title}
  {\bibinfo {title} {{Assembly of hard spheres in a cylinder: a computational
  and experimental study}},\ }\href {https://doi.org/10.1039/C7SM00316A}
  {\bibfield  {journal} {\bibinfo  {journal} {Soft Matter}\ }\textbf {\bibinfo
  {volume} {13}},\ \bibinfo {pages} {3296} (\bibinfo {year}
  {2017})}\BibitemShut {NoStop}%
\bibitem [{\citenamefont {Meagher}\ \emph {et~al.}(2015)\citenamefont
  {Meagher}, \citenamefont {García-Moreno}, \citenamefont {Banhart},
  \citenamefont {Mughal},\ and\ \citenamefont
  {Hutzler}}]{10.1016/j.colsurfa.2014.12.020}%
  \BibitemOpen
  \bibfield  {author} {\bibinfo {author} {\bibfnamefont {A.}~\bibnamefont
  {Meagher}}, \bibinfo {author} {\bibfnamefont {F.}~\bibnamefont
  {García-Moreno}}, \bibinfo {author} {\bibfnamefont {J.}~\bibnamefont
  {Banhart}}, \bibinfo {author} {\bibfnamefont {A.}~\bibnamefont {Mughal}},\
  and\ \bibinfo {author} {\bibfnamefont {S.}~\bibnamefont {Hutzler}},\
  }\bibfield  {title} {\bibinfo {title} {{An experimental study of columnar
  crystals using monodisperse microbubbles}},\ }\href
  {https://doi.org/10.1016/j.colsurfa.2014.12.020} {\bibfield  {journal}
  {\bibinfo  {journal} {Colloids and Surf. A}\ }\textbf {\bibinfo {volume}
  {473}},\ \bibinfo {pages} {55} (\bibinfo {year} {2015})}\BibitemShut
  {NoStop}%
\bibitem [{\citenamefont {Yamchi}\ and\ \citenamefont
  {Bowles}(2015)}]{10.1103/physrevlett.115.025702}%
  \BibitemOpen
  \bibfield  {author} {\bibinfo {author} {\bibfnamefont {M.~Z.}\ \bibnamefont
  {Yamchi}}\ and\ \bibinfo {author} {\bibfnamefont {R.~K.}\ \bibnamefont
  {Bowles}},\ }\bibfield  {title} {\bibinfo {title} {{Helical Defect Packings
  in a Quasi-One-Dimensional System of Cylindrically Confined Hard Spheres}},\
  }\href {https://doi.org/10.1103/physrevlett.115.025702} {\bibfield  {journal}
  {\bibinfo  {journal} {Phys. Rev. Lett.}\ }\textbf {\bibinfo {volume} {115}},\
  \bibinfo {pages} {025702} (\bibinfo {year} {2015})}\BibitemShut {NoStop}%
\bibitem [{\citenamefont {Torquato}\ and\ \citenamefont
  {Stillinger}(2001)}]{10.1021/jp011960q}%
  \BibitemOpen
  \bibfield  {author} {\bibinfo {author} {\bibfnamefont {S.}~\bibnamefont
  {Torquato}}\ and\ \bibinfo {author} {\bibfnamefont {F.~H.}\ \bibnamefont
  {Stillinger}},\ }\bibfield  {title} {\bibinfo {title} {{Multiplicity of
  Generation, Selection, and Classification Procedures for Jammed Hard-Particle
  Packings}},\ }\href {https://doi.org/10.1021/jp011960q} {\bibfield  {journal}
  {\bibinfo  {journal} {J. Phys. Chem. B}\ }\textbf {\bibinfo {volume} {105}},\
  \bibinfo {pages} {11849} (\bibinfo {year} {2001})}\BibitemShut {NoStop}%
\bibitem [{mv1(2020)}]{mv12}%
  \BibitemOpen
  \href@noop {} {\bibinfo {title} {{Wolfram Research Inc, Mathematica 12.2}}}
  (\bibinfo {year} {2020})\BibitemShut {NoStop}%
\bibitem [{sup(2021)}]{supmat}%
  \BibitemOpen
  \bibfield  {title} {\bibinfo {title} {{Supplemental Material, url, for
  Inherent Structure Landscape of Hard Spheres Confined to Narrow Cylindrical
  Channels.}},\ }\href@noop {} {\  (\bibinfo {year} {2021})}\BibitemShut
  {NoStop}%
\bibitem [{\citenamefont {Lubachevsky}\ and\ \citenamefont
  {Stillinger}(1990)}]{10.1007/bf01025983}%
  \BibitemOpen
  \bibfield  {author} {\bibinfo {author} {\bibfnamefont {B.~D.}\ \bibnamefont
  {Lubachevsky}}\ and\ \bibinfo {author} {\bibfnamefont {F.~H.}\ \bibnamefont
  {Stillinger}},\ }\bibfield  {title} {\bibinfo {title} {{Geometric properties
  of random disk packings}},\ }\href {https://doi.org/10.1007/bf01025983}
  {\bibfield  {journal} {\bibinfo  {journal} {J. Stat. Phys.}\ }\textbf
  {\bibinfo {volume} {60}},\ \bibinfo {pages} {561} (\bibinfo {year}
  {1990})}\BibitemShut {NoStop}%
\bibitem [{\citenamefont {Kurchan}\ \emph {et~al.}(2012)\citenamefont
  {Kurchan}, \citenamefont {Parisi},\ and\ \citenamefont
  {Zamponi}}]{10.1088/1742-5468/2012/10/p10012}%
  \BibitemOpen
  \bibfield  {author} {\bibinfo {author} {\bibfnamefont {J.}~\bibnamefont
  {Kurchan}}, \bibinfo {author} {\bibfnamefont {G.}~\bibnamefont {Parisi}},\
  and\ \bibinfo {author} {\bibfnamefont {F.}~\bibnamefont {Zamponi}},\
  }\bibfield  {title} {\bibinfo {title} {{Exact theory of dense amorphous hard
  spheres in high dimension I. The free energy}},\ }\href
  {https://doi.org/10.1088/1742-5468/2012/10/p10012} {\bibfield  {journal}
  {\bibinfo  {journal} {J. Stat. Mech.}\ }\textbf {\bibinfo {volume} {2012}},\
  \bibinfo {pages} {P10012} (\bibinfo {year} {2012})}\BibitemShut {NoStop}%
\bibitem [{\citenamefont {Kurchan}\ \emph {et~al.}(2013)\citenamefont
  {Kurchan}, \citenamefont {Parisi}, \citenamefont {Urbani},\ and\
  \citenamefont {Zamponi}}]{10.1021/jp402235d}%
  \BibitemOpen
  \bibfield  {author} {\bibinfo {author} {\bibfnamefont {J.}~\bibnamefont
  {Kurchan}}, \bibinfo {author} {\bibfnamefont {G.}~\bibnamefont {Parisi}},
  \bibinfo {author} {\bibfnamefont {P.}~\bibnamefont {Urbani}},\ and\ \bibinfo
  {author} {\bibfnamefont {F.}~\bibnamefont {Zamponi}},\ }\bibfield  {title}
  {\bibinfo {title} {{Exact Theory of Dense Amorphous Hard Spheres in High
  Dimension. II. The High Density Regime and the Gardner Transition}},\ }\href
  {https://doi.org/10.1021/jp402235d} {\bibfield  {journal} {\bibinfo
  {journal} {J. Phys. Chem. B}\ }\textbf {\bibinfo {volume} {117}},\ \bibinfo
  {pages} {12979} (\bibinfo {year} {2013})}\BibitemShut {NoStop}%
\bibitem [{\citenamefont {Charbonneau}\ \emph {et~al.}(2014)\citenamefont
  {Charbonneau}, \citenamefont {Kurchan}, \citenamefont {Parisi}, \citenamefont
  {Urbani},\ and\ \citenamefont {Zamponi}}]{10.1038/ncomms4725}%
  \BibitemOpen
  \bibfield  {author} {\bibinfo {author} {\bibfnamefont {P.}~\bibnamefont
  {Charbonneau}}, \bibinfo {author} {\bibfnamefont {J.}~\bibnamefont
  {Kurchan}}, \bibinfo {author} {\bibfnamefont {G.}~\bibnamefont {Parisi}},
  \bibinfo {author} {\bibfnamefont {P.}~\bibnamefont {Urbani}},\ and\ \bibinfo
  {author} {\bibfnamefont {F.}~\bibnamefont {Zamponi}},\ }\bibfield  {title}
  {\bibinfo {title} {{Fractal free energy landscapes in structural glasses}},\
  }\href {https://doi.org/10.1038/ncomms4725} {\bibfield  {journal} {\bibinfo
  {journal} {Nat. Commun.}\ }\textbf {\bibinfo {volume} {5}},\ \bibinfo {pages}
  {3725} (\bibinfo {year} {2014})}\BibitemShut {NoStop}%
\bibitem [{\citenamefont {Berthier}\ \emph {et~al.}(2019)\citenamefont
  {Berthier}, \citenamefont {Biroli}, \citenamefont {Charbonneau},
  \citenamefont {Corwin}, \citenamefont {Franz},\ and\ \citenamefont
  {Zamponi}}]{10.1063/1.5097175}%
  \BibitemOpen
  \bibfield  {author} {\bibinfo {author} {\bibfnamefont {L.}~\bibnamefont
  {Berthier}}, \bibinfo {author} {\bibfnamefont {G.}~\bibnamefont {Biroli}},
  \bibinfo {author} {\bibfnamefont {P.}~\bibnamefont {Charbonneau}}, \bibinfo
  {author} {\bibfnamefont {E.~I.}\ \bibnamefont {Corwin}}, \bibinfo {author}
  {\bibfnamefont {S.}~\bibnamefont {Franz}},\ and\ \bibinfo {author}
  {\bibfnamefont {F.}~\bibnamefont {Zamponi}},\ }\bibfield  {title} {\bibinfo
  {title} {{Gardner physics in amorphous solids and beyond}},\ }\href
  {https://doi.org/10.1063/1.5097175} {\bibfield  {journal} {\bibinfo
  {journal} {J. Chem. Phys.}\ }\textbf {\bibinfo {volume} {151}},\ \bibinfo
  {pages} {010901} (\bibinfo {year} {2019})}\BibitemShut {NoStop}%
\bibitem [{\citenamefont {Hove}(1950)}]{10.1016/0031-8914(50)90072-3}%
  \BibitemOpen
  \bibfield  {author} {\bibinfo {author} {\bibfnamefont {L.~v.}\ \bibnamefont
  {Hove}},\ }\bibfield  {title} {\bibinfo {title} {{Sur L'intégrale de
  Configuration Pour Les Systèmes De Particules À Une Dimension}},\ }\href
  {https://doi.org/10.1016/0031-8914(50)90072-3} {\bibfield  {journal}
  {\bibinfo  {journal} {Physica}\ }\textbf {\bibinfo {volume} {16}},\ \bibinfo
  {pages} {137} (\bibinfo {year} {1950})}\BibitemShut {NoStop}%
\bibitem [{\citenamefont {Lieb}\ and\ \citenamefont {Mattis}(1966)}]{lieb}%
  \BibitemOpen
  \bibfield  {author} {\bibinfo {author} {\bibfnamefont {E.}~\bibnamefont
  {Lieb}}\ and\ \bibinfo {author} {\bibfnamefont {D.}~\bibnamefont {Mattis}},\
  }\href@noop {} {\emph {\bibinfo {title} {{Mathematical Physics in One
  Dimension}}}}\ (\bibinfo  {publisher} {New York, NY, Academic Press},\
  \bibinfo {year} {1966})\BibitemShut {NoStop}%
\bibitem [{\citenamefont {Landau}\ and\ \citenamefont
  {Lifshitz}(1980)}]{landau}%
  \BibitemOpen
  \bibfield  {author} {\bibinfo {author} {\bibfnamefont {L.}~\bibnamefont
  {Landau}}\ and\ \bibinfo {author} {\bibfnamefont {E.}~\bibnamefont
  {Lifshitz}},\ }\href@noop {} {\emph {\bibinfo {title} {{Statistical Physics,
  Part 1. Course of Theoretical Physics Volume 5 }}}}\ (\bibinfo  {publisher}
  {Oxford ; New York : Pergamon Press},\ \bibinfo {year} {1980})\BibitemShut
  {NoStop}%
\bibitem [{\citenamefont {Thouless}(1969)}]{10.1103/PhysRev.187.732}%
  \BibitemOpen
  \bibfield  {author} {\bibinfo {author} {\bibfnamefont {D.~J.}\ \bibnamefont
  {Thouless}},\ }\bibfield  {title} {\bibinfo {title} {{Long-Range Order in
  One-Dimensional Ising Systems}},\ }\href
  {https://doi.org/10.1103/physrev.187.732} {\bibfield  {journal} {\bibinfo
  {journal} {Phys. Rev.}\ }\textbf {\bibinfo {volume} {187}},\ \bibinfo {pages}
  {732} (\bibinfo {year} {1969})}\BibitemShut {NoStop}%
\bibitem [{\citenamefont {Kosterlitz}\ and\ \citenamefont
  {Thouless}(2016)}]{10.1142/S0217979216300188}%
  \BibitemOpen
  \bibfield  {author} {\bibinfo {author} {\bibfnamefont {J.~M.}\ \bibnamefont
  {Kosterlitz}}\ and\ \bibinfo {author} {\bibfnamefont {D.~J.}\ \bibnamefont
  {Thouless}},\ }\bibfield  {title} {\bibinfo {title} {{Early Work on Defect
  Driven Phase Transitions}},\ }\href
  {https://doi.org/10.1142/s0217979216300188} {\bibfield  {journal} {\bibinfo
  {journal} {Int. J. Mod. Phys. B}\ }\textbf {\bibinfo {volume} {30}},\
  \bibinfo {pages} {1630018} (\bibinfo {year} {2016})}\BibitemShut {NoStop}%
\bibitem [{\citenamefont {Hu}\ \emph {et~al.}(2018)\citenamefont {Hu},
  \citenamefont {Fu},\ and\ \citenamefont
  {Charbonneau}}]{10.1080/00268976.2018.1479543}%
  \BibitemOpen
  \bibfield  {author} {\bibinfo {author} {\bibfnamefont {Y.}~\bibnamefont
  {Hu}}, \bibinfo {author} {\bibfnamefont {L.}~\bibnamefont {Fu}},\ and\
  \bibinfo {author} {\bibfnamefont {P.}~\bibnamefont {Charbonneau}},\
  }\bibfield  {title} {\bibinfo {title} {{Correlation lengths in
  quasi-one-dimensional systems via transfer matrices}},\ }\href
  {https://doi.org/10.1080/00268976.2018.1479543} {\bibfield  {journal}
  {\bibinfo  {journal} {Molecular Physics}\ }\textbf {\bibinfo {volume}
  {116}},\ \bibinfo {pages} {1} (\bibinfo {year} {2018})}\BibitemShut {NoStop}%
\bibitem [{\citenamefont {Yilmaz}\ and\ \citenamefont
  {Zimmermann}(2005)}]{10.1103/physreve.71.026127}%
  \BibitemOpen
  \bibfield  {author} {\bibinfo {author} {\bibfnamefont {M.~B.}\ \bibnamefont
  {Yilmaz}}\ and\ \bibinfo {author} {\bibfnamefont {F.~M.}\ \bibnamefont
  {Zimmermann}},\ }\bibfield  {title} {\bibinfo {title} {{Exact cluster size
  distribution in the one-dimensional Ising model}},\ }\href
  {https://doi.org/10.1103/physreve.71.026127} {\bibfield  {journal} {\bibinfo
  {journal} {Phys. Rev. E}\ }\textbf {\bibinfo {volume} {71}},\ \bibinfo
  {pages} {026127} (\bibinfo {year} {2005})}\BibitemShut {NoStop}%
\end{thebibliography}

%


\end{document}